\documentclass[jcp,aip,amsfonts,twocolumn,reprint]{revtex4-1} 
\usepackage{graphicx}
\usepackage{amsmath}
\usepackage{indentfirst}
\usepackage{epstopdf}
\usepackage[hyperindex,breaklinks]{hyperref} 

\begin{document}
\title{Reduced Dimension DVR Study of \emph{cis}-\emph{trans} Isomerization in the S$_{1}$ State of C$_{2}$H$_{2}$}
\author{J.~H.~Baraban$^a$, A.~R.~Beck$^b$, A.~H.~Steeves$^c$, J.~F.~Stanton$^d$, R.~W.~Field$^{a,*}$}
\noaffiliation
\affiliation{$^a$Department of Chemistry, Massachusetts Institute of Technology, Cambridge, Massachusetts 02139, USA\\
$^b$Department of Chemistry, University of California, Berkeley, California 94720\\
$^c$Department of Pharmaceutical Chemistry, University of California at San Francisco, San Francisco, CA 94143, USA\\
$^d$Institute for Theoretical Chemistry, Departments of Chemistry and Biochemistry, The University of Texas at Austin, Austin, Texas 78712\\
$^{*}$Author to whom correspondence should be addressed.  Tel (617) 253-1489.  FAX (617) 253-7030.  Electronic mail:  rwfield@mit.edu}

\begin{abstract}
	 Isomerization between the \emph{cis} and \emph{trans} conformers of the S$_{1}$ state of acetylene is studied using a reduced dimension DVR calculation.   Existing DVR techniques are combined with a high accuracy potential energy surface and a kinetic energy operator derived from $\mathbf{FG}$ theory to yield an effective but simple Hamiltonian for treating large amplitude motions.  The spectroscopic signatures of the S$_{1}$ isomerization are discussed, with emphasis on the vibrational aspects.  The presence of a low barrier to isomerization causes distortion of the \emph{trans} vibrational level structure and the appearance of nominally electronically forbidden $\tilde{A}\ {}^{1}A_{2}\leftarrow\tilde{X}\ {}^{1}\Sigma_{g}^{+}$ transitions to vibrational levels of the \emph{cis} conformer.  Both of these effects are modeled in agreement with experimental results, and the underlying mechanisms of tunneling and state mixing are elucidated by use of the calculated vibrational wavefunctions.  
\end{abstract}

\maketitle

\section{Introduction}

The history of the $\tilde{A}$ state of acetylene is full of spectroscopic surprises.  Beginning with the discovery that acetylene changes shape from linear to \emph{trans}-bent upon electronic excitation \cite{king_bent_1952,ingold1953esa,innes:863}, the $\tilde{A}\leftarrow\tilde{X}$ spectrum has exhibited a string of unexpected phenomena, including axis switching \cite{hougen1965anomalous}, triplet perturbations \cite{abramson:2293,Dupre1991293,drabbels:165}, as well as unusually strong Coriolis interactions and Darling-Dennison resonance between the bending vibrations \cite{utz:2742,merer:054304}.  Most recently, the comprehensive assignment of the low energy vibrational structure has led to the identification of several ``extra'' levels.   These appear in the spectrum near 3000 cm$^{-1}$ above the ground vibrational state of the {\it trans} conformer, just as the effective Hamiltonian ($\mathbf{H}^{\textrm{eff}}$) models developed at lower energy begin to break down \cite{steeves2009stretchbend}.

The most probable explanation for these ``extra'' levels is that they belong to the \emph{cis} conformer of the S$_{1}$ state \cite{46175}.  \emph{Ab initio} calculations have long predicted a \emph{cis} minimum \cite{Demoulin1975329,stanton:356,ventura2003ves}, but experimental confirmation has been hard to come by.  This is because the S$_{1}$-S$_{0}$ electronic transition is forbidden in $C_{2v}$ geometries, and any spectroscopic observation of the \emph{cis} conformer is therefore expected to manifest itself in the form of weak transitions or indirect effects, necessarily resulting from vibronic mechanisms.


The purpose of this paper is to investigate theoretically the spectroscopic consequences of the low barrier isomerization process on the $\tilde{A}$ state potential surface.  In general, how will the presence of a second distinct accessible minimum distort the vibrational energy level patterns?  One detail of particular interest is whether S$_{1}$ \emph{cis} states could appear at the observed energies with the intensities observed for the ``extra'' levels.  The calculations in the literature characterize the relevant stationary points on the potential energy surface (PES), but cannot address these questions because of the delocalized nature of the wavefunctions involved in large amplitude motion and the importance of effects like tunneling.   Large amplitude dynamics on a reduced dimension S$_{1}$ surface have been studied previously by spectral quantization \cite{2001PCCP....3.5393M,Schubert:2005fk}, but those calculations focused mainly on the well-known Franck-Condon active progressions in the CC stretching mode, $v_{2}$, and the {\it trans}-bending mode, $v_{3}$, while revealing little about the weak or forbidden bands that encode the isomerization dynamics.

The method we choose for our treatment uses a reduced dimension Discrete Variable Representation \cite{light2000dvr}.  Since we are investigating an isomerization process, the calculation must be able to treat large amplitude motions that span the two geometries.  DVR methods are well-suited to this type of problem because the basis functions are not tied to a single center as they usually are in a variational calculation \cite{bacic:4594}.  Since the half-linear transition state \cite{stanton:356} and the Franck-Condon active vibrations are all planar, we can perform a reduced dimension calculation and still expect to capture the essential features of the experimental spectrum.  As described in detail below, the calculation is performed in the three dimensions that encompass the CC stretch as well as the \emph{trans} and \emph{cis} bending vibrations.

An important ingredient in our calculation is a high accuracy potential energy surface for an excited electronic state.  This work represents the first application of EOM-CCSDT \cite{2001JChPh.115.8263K} methods to the PES of a polyatomic molecule, and we believe there is great promise in applying such methods to the study of electronically excited states.  In fact, we find that this level of theory is necessary to obtain quantitatively acceptable agreement with experiment for this system.

We obtain from our DVR calculation the vibrational eigenstates of the full S$_{1}$ potential surface in the three dimensional coordinate space.  Since the basis functions are not defined relative to one specific geometry, the eigenfunctions are not predetermined to belong to either the \emph{cis} or \emph{trans} conformer, although they naturally divide themselves in this way at low energy.  We therefore expect that the calculated results will contain all of the effects we desire to model: the possible existence of \emph{cis} levels interspersed among those of the \emph{trans} conformer, and any mixings between them that arise from tunneling through the isomerization barrier.  Other signatures of the isomerization, such as the distortion of the \emph{trans} level structure, should also be satisfactorily reproduced.

Finally, we note that this method could be generally applicable to other isomerizations, even in larger molecules.  It allows for a selective treatment of only the few coordinates relevant to the minimum energy isomerization path, but is still based on a simple and easily understandable Hamiltonian.  Its foundation on established DVR methods provides computational efficiency and accuracy.

\section{Methods}\label{sec:methods}

In this section we describe the elements of our method for reduced dimension vibrational DVR calculations.  Ours is not the first such calculation, even on acetylene  \cite{bentley:4255,sibert:937}, 
but our approach differs from those of previous authors.  We specify significant departures from earlier reduced dimension DVR calculations on acetylene where appropriate.

An important consideration in a reduced dimension calculation is that the qualitative and quantitative accuracy of the results are frequently balanced against the effort required to perform the calculation.  In this work we have gone to some lengths in order to obtain quantitatively accurate results, but very useful qualitative results can be had much more easily.  Even for spectroscopists, the pattern and assignments of the levels can be valuable even if quantitative agreement with experiment is poor.  We demonstrate this point later on in connection with the quality of the potential energy surface and the number of dimensions included in the calculation, which are the two factors that have the greatest impact on the computational cost of the calculation.

It is appropriate to begin our discussion of the details by mentioning that we desire a matrix representation of $H=T + V$, where $T$ and $V$ are expressed in the same basis.  We will examine the two parts of the Hamiltonian separately, but first we must digress briefly on the subject of coordinate systems.

\subsection{Molecule-Fixed Coordinate Systems}

The choice of coordinate system for a calculation in the molecular frame is a complicated one.  The problem of separating vibration and rotation is exacerbated when large amplitude motions are possible  \cite{bunker2006msa}, and even remaining in the center of mass frame may not be trivial  \cite{bentley:4255,sibert:937} for molecules larger than triatomics.  For acetylene, several different coordinate systems have been used in the literature for various types of calculations  \cite{prosmiti:3299,bramley1991vrc}.  We find the internal coordinates of $\mathbf{FG}$ theory  \cite{wilson1980mvt} to be extremely convenient, and recommend them generally.  The potential energy surface is compactly represented in these coordinates, especially at low energy, and the $\mathbf{G}$ matrix elements  \cite{decius:1025,ferigle:982} derived from them provide a simple way to represent the kinetic energy operator.  

In reduced dimension calculations using internal coordinates, one departure that may be necessary from the canonical set of such coordinates for a given molecule concerns their domains.  Since the usual set of $3N-6$ coordinates is chosen to uniquely specify all possible geometries (leaving aside linearity for the moment), it may be necessary to change the domain of one or more of the angles.  For example, since we do not include the usual torsional coordinate, $\tau$, in our planar acetylene calculation, it is necessary to increase the domain maximum of at least one of the $\angle$CCH valence angle bends from $\pi$ to $2\pi$ in order to allow both \textit{cis} and \textit{trans} geometries.  (In this paper $\tau$ is defined as zero, except in Table \ref{tab:HOfreqs} where $\tau=\pi$.)

\subsection{The Potential Energy Surface}
\label{sec:PES}
The hallmark of all DVR calculations is the simplicity of the potential energy matrix elements.  More generally, any function of coordinates is a diagonal matrix, $\langle \theta_{i}|\hat{f}(x)|\theta_{j}\rangle\approx f(x_{i})\delta_{ij}$, where the  \{$x_{i}$\} are the pseudospectral grid points of the DVR basis \cite{tannor2007iqm}.  Once a reduced set of coordinates has been chosen, only two things remain to be done to express $V$ in the DVR basis: finding the appropriate level of \textit{ab initio} theory at which to calculate the potential energy surface, and deciding how to treat the discarded degrees of freedom when the active coordinates are varied.

Previous theoretical investigations of the S$_{1}$ surface encountered difficulties with the harmonic frequencies of the \textit{trans} structure, especially the lowest frequency modes, $v_{4}$ and $v_{6}$ \cite{stanton:356,ventura2003ves}.  We report here EOM-CCSDT harmonic frequencies and compare them to EOM-CCSD values and experimental data in Table \ref{tab:HOfreqs}.  Since we exclude $v_{4}$, the CCSD frequencies seem satisfactory, but still the improved agreement at CCSDT is obvious.  Indeed, we were able to obtain qualitatively useful results with a CCSD potential surface, but quantitative agreement requires CCSDT\@.  In light of this, a reasonable full set of harmonic frequencies appears to be a good criterion for selecting a level of \emph{ab initio} theory.  The relatively poor quantitative agreement with experiment obtained in a full dimensional calculation of the vibrational fundamentals using a CCSD surface \cite{rheinecker2006aic} corroborates this conclusion, as our reduced dimension CCSD vibrational fundamentals for $v_{3}$ and $v_{6}$ are very similar to their results.

\vbox{
\makeatletter
\def\@captype{figure}
\begin{center}
\includegraphics[width=\linewidth]{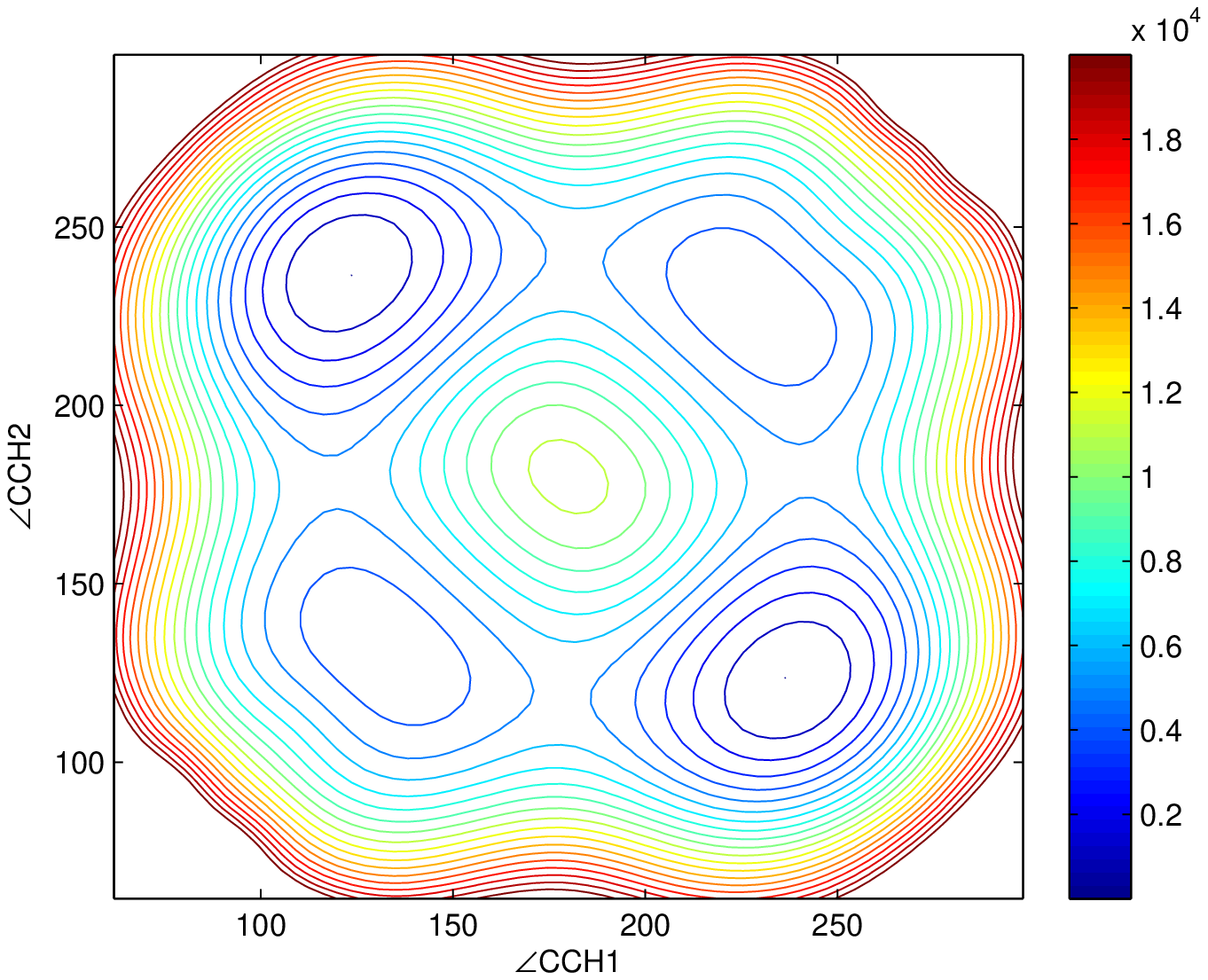}
\end{center}
\caption{\footnotesize
\label{fig:S1surf}
A slice of the S$_{1}$ PES with $R_{\textrm{CC}}$ and $R_{\textrm{CH}}$ at their \emph{trans} equilibrium values.  The \emph{cis} minima are located in the lower left and upper right, and the \emph{trans} minima are in the upper left and lower right. The energy scale is in cm$^{-1}$ and the contour spacing is $10^{3}$ cm$^{-1}$.
	}
}  

\begin{table}
\caption{Geometries and harmonic frequencies for $\tilde{A}\ ^{1}A_{u}$ $^{12}$C$_{2}$H$_{2}$: Comparison of \emph{ab initio} methods }
\vspace{8pt}
\begin{tabular*}{0.85\linewidth}{@{\extracolsep{\fill}} c | c c c}
&CCSD \cite{stanton:356} & CCSDT & Expt. \cite{merer:054304,steeves:1867,tobiason:5762,Huet199032}\\
\hline
$\omega_{1}$ & 3107.7 & 3053.5 & (3004)\footnotemark[1]\\
$\omega_{2}$ & 1471.4 & 1424.6 & 1410.26\\
$\omega_{3}$ & 1106.6 & 1106.1 & 1070.34\\
$\omega_{4}$ & 614.6 & 761.5 & 764.7\\
$\omega_{5}$ & 3083.7 & 3031.0 & (2914)\footnotemark[1]\\
$\omega_{6}$ & 745.8 & 776.5 & 772.5\\
\hline
$\angle$CCH \ &  123.64 & 122.09 & 122.48\\
$R_{\textrm{CC}}$ & 1.3575 & 1.3743 & 1.375\\
$R_{\textrm{CH}}$ & 1.0907 & 1.0963 & 1.097\\
$T_{0}$ & 43830 & 42334 & 42197.57\\
\end{tabular*}
\footnotetext[0]{Frequencies in cm$^{-1}$, angles in degrees, and bond lengths in \AA.}
\footnotetext[1]{These harmonic frequencies are less well determined because their overtones have not yet been observed. \cite{steeves:1867}}
\label{tab:HOfreqs}
\end{table}
The simplest approach for dealing with the discarded degrees of freedom is to fix them at their equilibrium values.  While it is impossible to make absolute generalizations due to the heterogeneous decomposition of the normal modes into the internal coordinates, our experience is that this approximation leads to computed vibrational fundamentals within at worst 10\% of the experimental values, but usually significantly better.  We would have been satisfied with this approximation except for the Fermi resonance that manifested itself between $2^{n}6^{m}$ and $2^{n-1}6^{m+2}$ in early calculated results, which precluded comparisons with effective Hamiltonian fits to experimental data.  We therefore replace $R_{\textrm{CC}}$ with the normal coordinate for $v_{2}$ from the harmonic frequency calculation, which provides a prescription for varying the CH distances as the CC bond length changes.   \cite{footnote1}  We did also consider vibrational adiabatic potentials, which are commonly used in the literature \cite{sibert:937,bentley:4255}.  However, given the limited return and the enormous computational cost of minimizing the energy at every point on the entire three dimensional grid, we elected not to pursue this option.  A fitted surface might be more suitable for addressing this particular issue.

\subsection{The Kinetic Energy Operator}
\label{sec:KE}
The kinetic energy operator presents several difficulties.  The form of the Laplacian in the chosen coordinate system must be determined, which can be very tricky \cite{bramley1991vrc,bentley:4255,sibert:937}, especially since singularities in the kinetic energy are possible. Furthermore, expressing the Laplacian in a DVR basis can lead to a non-Hermitian $\mathbf{H}$\@.  The kinetic energy operator matrix elements are more complicated than those of $V$, but we merely need to add matrix representations of differential operators to the trivial matrix representations of functions of coordinates already in our repertoire.  Depending on the underlying basis, DVR matrix elements of differential operators can be calculated straightforwardly either via basis transformations to and from a finite basis representation \cite{harris1965cme}, e.g.~a harmonic oscillator, or by explicit formulas \cite{szalay:1978}, most commonly based on Fourier functions \cite{marston:3571,colbert:1982,tuvi:9079}.  The selection of an appropriate DVR basis consists primarily of choosing one whose underlying basis functions have the same boundary conditions as the eigenfunctions to be obtained \cite{colbert:1982}.

We use the $\mathbf{G}$ matrix elements to obtain the kinetic energy operator, taking care to follow Podolsky \cite{podolsky}, and also using an explicitly Hermitian form \cite{wei:1343} of each term, such that
\begin{equation}
T=\dfrac{-\hbar^{2}}{2}g^{1/4}\sum_{i,j}\left(\dfrac{\partial}{\partial S_{i}}\right)^{\dag}g^{-1/2}\mathbf{G}_{ij}\dfrac{\partial}{\partial S_{j}}g^{1/4}
\label{eq:KE}\end{equation}
where $g=\textrm{det}\,|\mathbf{G}|$ and the $S_{i}$ are the internal coordinates.  It is important to note that although the traditional $\mathbf{FG}$ matrix solution invokes the approximation of infinitesimal displacements, the coordinates themselves and their $\mathbf{G}$ matrix elements are valid for motions of any amplitude \cite{wilson1980mvt}.  We simply represent the functions of coordinates in the $\mathbf{G}_{ij}$ as diagonal matrices instead of evaluating them at a specific geometry.  In our planar acetylene calculation we take care to use the kinetic energy matrix elements for the bending of a linear molecule  \cite{ferigle:982}, since the torsional angle is not well behaved when planar \textit{cis}-\textit{trans} isomerization can occur.

Finally, for consistency with our use of the $v_{2}$ normal coordinate when calculating the PES, we also redefine $R_{\textrm{CC}}$ as the CC stretch component of the $v_{2}$ normal coordinate, according to $Q_{k}=\sum_{t}\left(L^{-1}\right)_{kt}S_{t}$, where $Q$, $S$, and $L$ are the mass-weighted normal coordinate, an internal coordinate, and the transformation that diagonalizes $\mathbf{GF}$, respectively \cite{wilson1980mvt}${}^{,}$ \cite{footnote2}.
The $R_{\textrm{CH}}$ contributions are still neglected.

\subsection{Specifics and Practical Considerations}
The specifics of our treatment of S$_{1}$ acetylene are as follows.  We take a subset of the internal coordinates, consisting of the two $\angle$CCH valence angle bends ($0\leq \{\phi_{1},\phi_{2}\}\leq2\pi$) and the C--C bond stretch, which correspond loosely to \{$v_{2},v_{3},v_{6}$\}.  Our DVR basis is then the direct product basis $\phi_{1}\otimes\phi_{2}\otimes R_{\textrm{CC}}$, and we use the sinc DVR basis \cite{colbert:1982} for the angles, but a harmonic oscillator basis for the bond stretch.  The final results were generated using 120 grid points for each of the two bend bases, and 43 points for the stretch basis.  The pseudospectral grid points and differential operator matrix elements are available from analytical formulas for the sinc basis, and, for the harmonic oscillator basis, from the diagonalization of the coordinate matrix and the application of the resulting transformation matrix to other operators expressed in the harmonic oscillator basis.  We then construct $\mathbf{T}$ according to Eq.~\ref{eq:KE}.   We calculate the potential energy for 10 geometries along each coordinate at regular intervals spanning \{$60^{\circ}\leq\phi_{1}\leq300^{\circ},60^{\circ}\leq\phi_{2}\leq180^{\circ}, 1$ \AA\ $\leq R_{\textrm{CC}}\leq2$ \AA \}. \emph{Ab initio} calculations were performed with the CFOUR program system \cite{cfour,mrcc}, using the EOM-CCSDT method and the NASA Ames ANO1 basis set.  The elements of $\mathbf{V}$ are found by interpolating the potential energy surface at the DVR grid points, after discarding grid points that lie outside the original domain of the \emph{ab initio} surface.  This reduced the size of the basis mentioned above to 80$\times$80$\times$34.

In practice, the construction and diagonalization of $\mathbf{H}$ are accomplished by a basis set contraction \cite{bacic:4594}.  The DVR $\mathbf{H}$ for the 2D $\phi_{1}\otimes\phi_{2}$ space at every value of $R_{\textrm{CC}}$ is diagonalized, and the eigenvectors above a chosen cutoff energy are discarded.  (For the results presented here this cutoff energy was 10,000 cm$^{-1}$, with a minimum of 35 vectors retained per 2D block.)  The resulting rectangular transformation matrices are used to compress the remaining blocks of the 3D $\mathbf{H}$, which are off-diagonal in $R_{\textrm{CC}}$, prior to the final diagonalization.  Applying the rectangular transformation matrices in reverse transforms the final eigenvectors back to the grid point representation.

\section{Results}

  Using the method described here, vibrational eigenstates of S$_{1}$ acetylene could be obtained up to energies exceeding 15,000 cm$^{-1}$.  For the most part we will limit the discussion to states up to about 5,000 cm$^{-1}$.  In general we will forego itemized examination of the calculated states and instead focus on more broad agreement with experiment and predicted trends, in keeping with the goals stated in the beginning of Section \ref{sec:methods}.  Nevertheless, it is worth mentioning that the calculated vibrational fundamentals are within 2\% of the experimental values (Table \ref{tab:vibfunds}), despite the neglect of the other dimensions.  (The results of an \emph{ab initio} harmonic frequency calculation for the \emph{cis} geometry are given for reference in Table \ref{tab:cisabinitio}.)

\begin{table}
\caption{DVR vibrational fundamentals for $\tilde{A}\ ^{1}A_{u}$ and $\tilde{A}\ ^{1}A_{2}$ $^{12}$C$_{2}$H$_{2}$}
\vspace{8pt}
\begin{tabular*}{.7\linewidth}{c | c c | c c}
&\multicolumn{2}{c|}{$\tilde{A}\ ^{1}A_{u}$}&\multicolumn{2}{c}{$\tilde{A}\ ^{1}A_{2}$}\\
               &Calc.       &Expt. \cite{utz:2742,Watson1982101} &Calc.&Expt. \cite{46175}\\ \hline
$v_{2}$ & 1414.65 &  1386.9 & 1489.61 & --\\
$v_{3}$ & 1033.6  &  1047.55 & 789.56 & --\\
$v_{6}$ & 780.35  & 768.26 & 588.35 & (565)\footnotemark[1]\\
\hline
\end{tabular*}
\footnotetext[0]{Frequencies in cm$^{-1}$.}
\footnotetext[1]{Estimated from the energy difference between two observed combination bands.}
\label{tab:vibfunds}
\end{table}

\begin{table}
\caption{\emph{Ab initio} geometry and harmonic frequencies for $\tilde{A}\ ^{1}A_{2}$ $^{12}$C$_{2}$H$_{2}$}
\vspace{8pt}
\begin{tabular*}{0.3\linewidth}{ c | c }
$\omega_{1}$ &\  2997.14\\
$\omega_{2}$ & 1583.22\\
$\omega_{3}$ & 806.09\\
$\omega_{4}$ & 817.5 \\
$\omega_{5}$ & 2941.81\\
$\omega_{6}$ & 571.62\\
\hline
$\angle$CCH \ &  132.62\\
$R_{\textrm{CC}}$ &  1.3423\\
$R_{\textrm{CH}}$ &  1.0983\\
$T_{0}$ & 45155\\
\end{tabular*}
\footnotetext[0]{Frequencies in cm$^{-1}$, angle in degrees, and bond lengths in \AA.}
\label{tab:cisabinitio}
\end{table}

\subsection{Symmetry of the Reduced Dimension Eigenstates}

	Before entering into a more detailed discussion of the calculated states, it is necessary to work out the connections between the symmetries that exist in the coordinate space of the calculation and the true symmetries of the molecule, so as to be able to interpret the wavefunctions.  It can be seen in Fig.~\ref{fig:S1surf} that two \emph{trans} and two \emph{cis} minima exist in the calculation.  Each well possesses a twofold symmetry, such that the PES has four equivalent (triangular) quadrants.  More generally, the symmetry group of the reduced dimension Hamiltonian is of order four.  Reflection across the antidiagonal corresponds generally to the $(12)(ab)$ CNPI symmetry operation \cite{bunker2006msa}, but can also be thought of as the $C_{2}^{b}$ operation in the $C_{2v}$ point group.  Reflection across the diagonal is similarly $(12)(ab)^{*}$, or the $i$ operation in the $C_{2h}$ point group.  This leads to the conclusion that inversion through the center of the coordinate space as depicted in Fig.~\ref{fig:S1surf} correlates with the $E^{*}$ operation.  Therefore, the eigenstates produced by the calculation will belong to one of the four rovibrational irreducible representations laid out in Table~\ref{tab:irreps}, as will be illustrated in Fig.~\ref{fig:4syms}.  Accordingly, we expect each calculated vibrational level to appear as a near degenerate pair of eigenstates with different rotational symmetries, except when \emph{cis}-\emph{trans} interaction causes one member of the pair with a particular rovibrational symmetry to shift.  This shift is analogous to the $K$-staggering that occurs in the experimental spectrum \cite{46175}.
	
\begin{table}
\caption{Rovibrational symmetries of DVR wavefunctions}
\vspace{8pt}
\begin{tabular*}{0.55\linewidth}{ c | c | c }
CNPI-MS \cite{lundberg1995CNPI} & $C_{2h}$ & $C_{2v}$\\
\hline
$Sa-/As-$ &  $a_{g},\ K_{c}$ odd\ & $\ b_{2}\ oo$\\
$Ss+/Aa+$ &  $\ a_{g},\ K_{c}$ even\ & $\ a_{1}\ ee$\\
$Ss-/Aa-$ &   $b_{u},\ K_{c}$ odd\ & $\ a_{1}\ oo$\\
$Sa+/As+$ & $\ b_{u},\ K_{c}$ even\ & $\ b_{2}\ ee$\\
\hline
\end{tabular*}
\footnotetext[0]{$R_{a}^{\pi}$ is not an equivalent rotation for the \emph{trans} geometry, and therefore its $oe/ee$ and $oo/eo$ have the same CNPI symmetry.  Conversely, the \emph{cis} rotational structure supports all CNPI symmetries, but only $oo$ and $ee$ appear here. The first member of a pair of CNPI-MS labels will be used as a shorthand notation.}
\label{tab:irreps}
\end{table}

\vbox{
\makeatletter
\def\@captype{figure}
\begin{center}
\includegraphics[width=\linewidth]{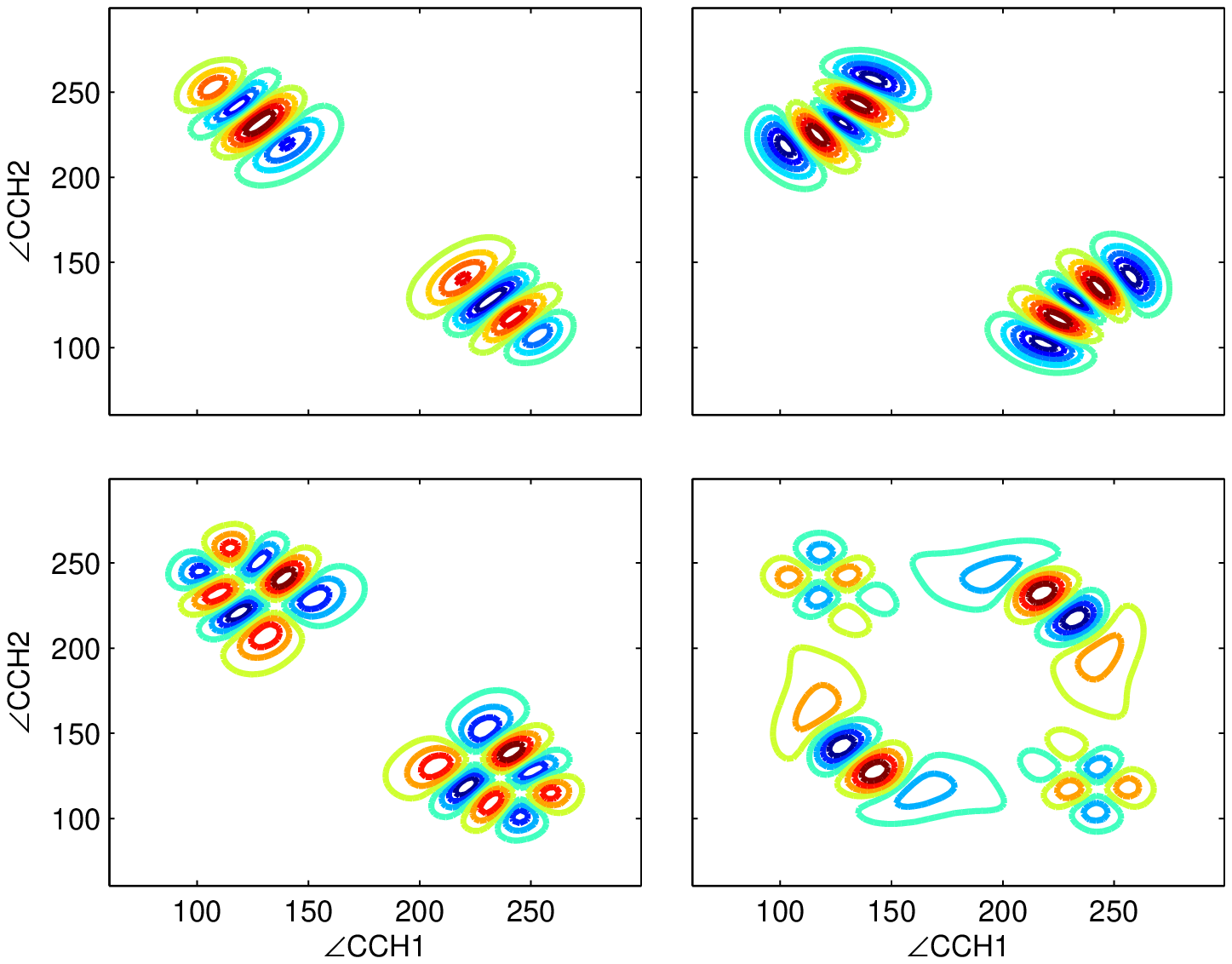}
\end{center}
\caption{\footnotesize
Wavefunctions illustrating the 4 symmetries listed in Table \ref{tab:irreps} and the types of levels discussed in Sec.~\ref{sec:agreement}-C.  Clockwise from the top left:  \emph{trans} $3^{3}$, $Sa-$; \emph{trans} $6^{4}$, $Ss+$; \emph{trans} $3^{3}6^{1}$, $Ss-$; \emph{cis} $6^{3}$, $Sa+$.  These states belong to the classes of levels with excitation in only totally symmetric modes, pure bending levels, stretch-bend combination levels, and \emph{cis} levels, respectively.  Note that the {\it cis} $6^{3}$ wavefunction is delocalized over both {\it cis} and {\it trans} wells.  The nodal patterns for \emph{trans} $v_{3}$ are similar to those obtained by spectral quantization \cite{2001PCCP....3.5393M,Schubert:2005fk}.
	}
	\label{fig:4syms}
}  

\subsection{Agreement with Experiment}
\label{sec:agreement}
	The primary difference between the DVR results and the observed levels is that many states are missing due to the reduced dimensionality.  Most important, due to the exclusion of $v_{4}$, all bending ($B$) polyads are represented by a single vibrational level, i.e.~instead of the $n+1$ vibrational levels in a $B^n$ polyad, the 3D calculation includes only one level, $6^{n}$.  States involving quanta in $v_{1}$ and $v_{5}$ are also missing, but less conspicuously. 
	
	The simplest illustration of the agreement between calculation and experiment is the juxtaposition of one photon spectra in Fig.~\ref{fig:1phcomp}.  (The method for computing spectra is described in Appendix A.)  The correspondence between individual features is clear at low vibrational energy, not only for the strong Franck-Condon active progressions that can be seen easily in the figure, but also for the weaker bending polyads \cite{steeves2009stretchbend,merer:054304} that have been identified. The comparison does become slightly more complicated at higher energy where the experimental characterization of the level structure is less complete.  The onset of predissociation \cite{mordaunt:2630} above 46074 cm$^{-1}$ causes many states to essentially disappear from the LIF spectrum, but the simulated spectrum makes no allowance for this effect.  Furthermore, the density of states not present in the reduced dimension calculation increases rapidly above the fundamentals of the neglected CH stretching modes (45077.65 and 45054.97 cm$^{-1}$ for $v_{1}$ and $v_{5}$, respectively).  Fortunately, these discrepancies do not cause significant difficulties in analyzing the calculated levels and extracting information relevant to the states observed experimentally.
	
	

\vbox{
\makeatletter
\def\@captype{figure}
\vspace{8pt}
\begin{center}
\includegraphics[width=\linewidth]{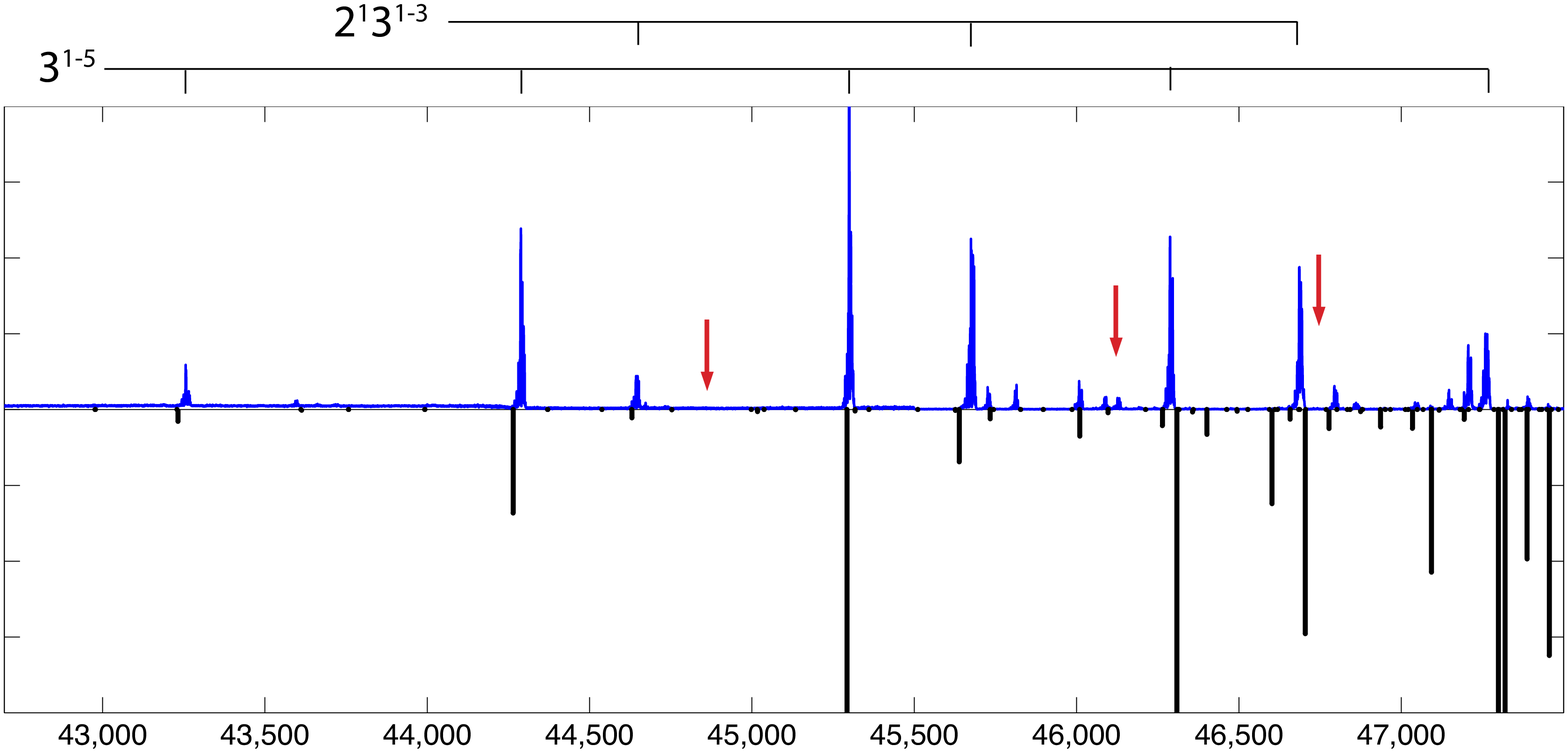}
\end{center}
\vspace{-16pt}
\caption{\footnotesize
Comparison of calculated spectrum (downward) with the one photon LIF \cite{steeves2009stretchbend}.  The more intense peaks have been truncated to make the weaker transitions visible.  Members of the two main Franck-Condon active progressions are identified for reference.  Three arrows mark the approximate positions of (from left to right) the {\it cis} origin, the C$_{2}$H+H dissociation limit, and the zero point energy corrected {\it cis}-{\it trans} barrier.
	}
	\label{fig:1phcomp}
}  

	We now proceed with a more detailed comparison of the calculated level structure to that observed experimentally.  In the recent experimental literature, the vibrational states are categorized in three groups: states containing excitation only in the totally symmetric modes \cite{steeves:1867,merer2003new}, the pure bending polyads \cite{merer:054304}, and the stretch-bend combination polyads \cite{steeves2009stretchbend}.  We organize our discussion accordingly.

\subsubsection{Totally Symmetric Modes}

	The states $2^{n}3^{m}$ are relatively straightforward, and the calculated levels match well with experiment.  Values obtained from fitting the groups $2^{n}$ and $3^{n}$ agree with Table 5 of Ref.~[\onlinecite{steeves:1867}], with $\omega_{2}$ and $\omega_{3}$ approximately 20 cm$^{-1}$ too high and too low, respectively.  The parameters $x_{22}$ and $x_{33}$ are close to correct but too small in magnitude, such that the residual for $v_{3}$ changes sign at $3v_{3}$.  Even for $v_{2}$, the residuals remain in the tens of cm$^{-1}$ over the energy region of interest.  The deviations from experiment are presumably due primarily to the neglect of the symmetric CH stretch.

\subsubsection{Pure Bending States}

	We next consider the pure bending polyads \cite{merer:054304}, where the lowest members are nominally the $6^{n}$ states that exist in the DVR calculation.  Here the comparison is complicated by the omission of the torsion from the calculation, since $v_{6}$ and $v_{4}$ interact very strongly via Coriolis effects and Darling-Dennison resonance.  Nevertheless, we find that $\omega_{6}$ from the DVR is only 8 cm$^{-1}$ higher than the experimental value, and the anharmonicity is once again too small in magnitude.  This is undoubtedly due to the surprisingly large contribution of the vibrational angular momentum to the experimental value of $x_{66}$, an effect absent from the calculation.  Overall the bending behavior here is nearly harmonic, as is the case with the deperturbed level structure.  As in the previous section, the too high harmonic frequency is presumably due at least in part to the neglect of the CH stretches, since $v_{5}$ is also of $b_{u}$ symmetry.

\subsubsection{Stretch-Bend Combination Polyads}

\begin{figure}
\includegraphics[width=\linewidth]{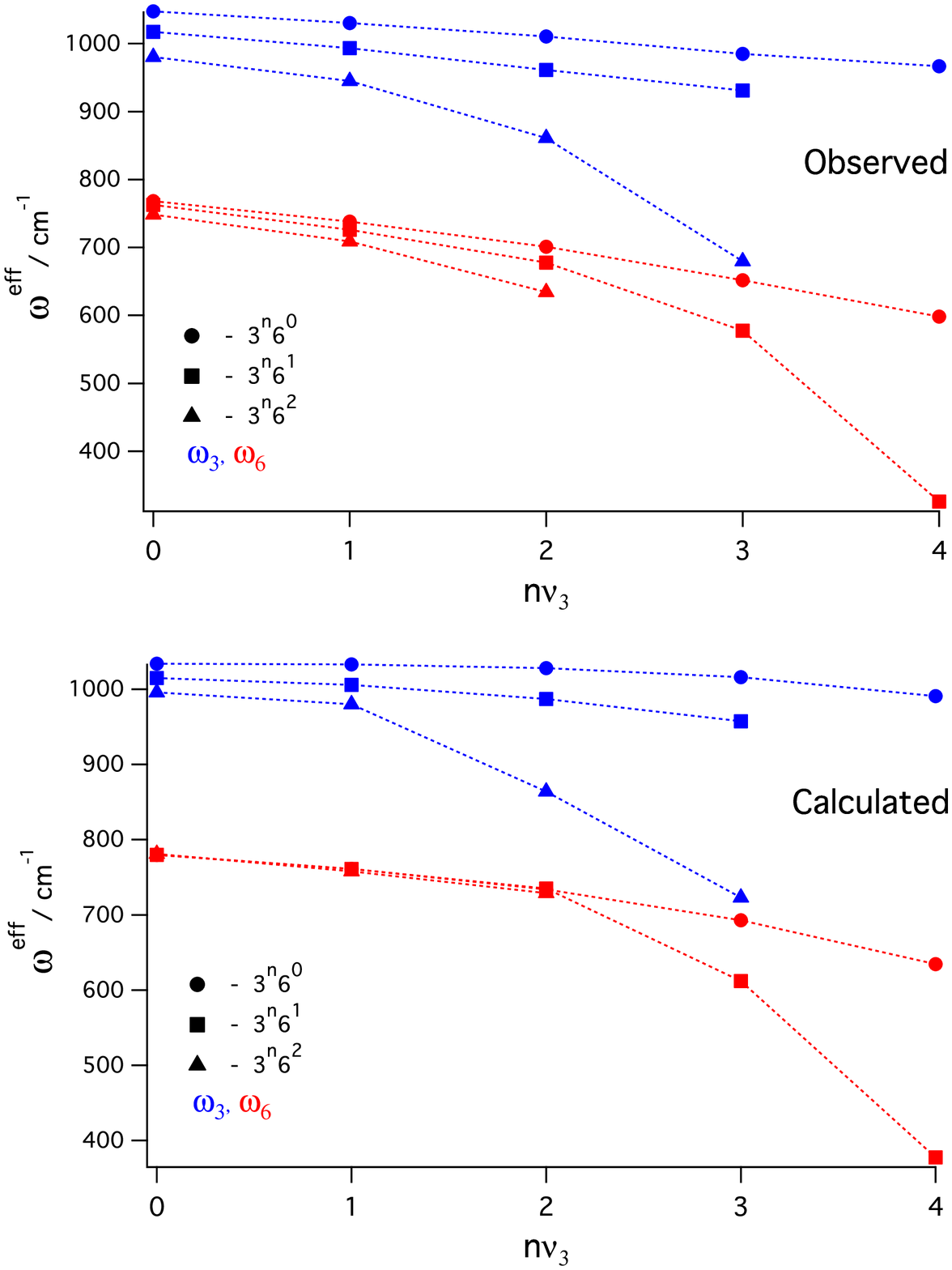}
\caption{
Plots of $\omega_{3}^{\textrm{eff}}$ and $\omega_{6}^{\textrm{eff}}$ vs.~$v_{3}$ derived from the observed and calculated level structure.  Different marker styles denote data from different values of $v_{6}$.  Deperturbed values for $T_{0}$ are used, except for the $3^{3}6^{2}$ and $3^{4}6^{2}$ levels, which have not yet been deperturbed. (In those cases the energy of the observed $J=K=0$ state is used instead.)  
}
	\label{fig:weffplots}
\end{figure}

	The stretch-bend combination polyads demand a more detailed comparison, because a global model that accounts for their vibrational structure has not yet been developed, despite the existence of extensive assignments and rotational analyses \cite{steeves2009stretchbend}.  Such a comparison is presented graphically in Fig.~\ref{fig:weffplots} by plotting the effective frequencies of the bending modes as a function of quanta in $v_{3}$.  It can be seen that the calculation reproduces well even the more unusual features of the observed level structure, and the minor differences are due to the disagreement in the diagonal anharmonicities, explained in the previous two sections.  The pathological behavior of this set of states is not entirely unexpected, as a combination of $v_{3}$ and $v_{6}$ essentially constitutes the isomerization path coordinate.  Excitation in both these modes in either well should promote \emph{cis}-\emph{trans} tunneling, which will cause mixings as discussed in subsequent sections.
	
	One oddity of particular interest is the dramatic decrease in $\omega_{3}^{\textrm{eff}}$ for the $3^{n}6^{2}$ series, illustrated in Fig.~13 of Ref.~\onlinecite{steeves2009stretchbend} up to $3^{2}6^{2}$.  Our recent high sensitivity spectra have revealed a band that is a promising candidate for $3^{3}6^{2}$, although the assignment has yet to be confirmed by rotational analysis and therefore details regarding it will be communicated later.  We nevertheless incorporate it in the figure for comparison with the DVR prediction.  We further include the well-known level at 47206 cm$^{-1}$ as $3^{4}6^{2}$, based on the preliminary assignment of $3^{3}6^{2}$ as well as previous discussions \cite{steeves:1867,steevesthesis}, in addition to the strong intensity and proximity of $3^{4}6^{2}$ to $3^{5}$ in the calculated spectra.  For both of these tentative assignments we find good agreement with the DVR results.  
	
	The importance of this sharp decrease in the effective frequencies is that it signals the onset of the \emph{cis}-\emph{trans} isomerization process as the potential softens approaching the transition state.  We now turn our attention to the \emph{cis} minimum of the $S_{1}$ state and its vibrational levels.
	

\subsection{$^{1}A_{2}$ \emph{cis} states}
\label{sec:cis}

The adiabatic energy separation between the \emph{cis} and \emph{trans} isomers
of acetylene is found to be 2664 cm$^{-1}$ based
on high-level {\it ab initio} calculations.   Specifically, the
effects of valence electron correlation have been included up to
the coupled-cluster singles, doubles, triples and quadruples (CCSDTQ)
level of theory, using the equation-of-motion
(EOM) variant of coupled-cluster theory to treat these excited state
isomers.   In addition, effects of basis set insufficiency
are estimated using extrapolation techniques \cite{tajti:11599},
and contributions due to zero point energies,
core correlation and scalar relativistic effects are included as
well.   The final result of these calculations predicts
the zero-point level of the \emph{cis} isomer to lie at 44861 cm$^{-1}$, a calculation that we believe
to be in error by no more than 50 cm$^{-1}$.   In passing, we note that the
present results are in line with a similar estimate
published some time ago by K\'{a}llay and Gauss \cite{kallay:9257} of 44852 cm$^{-1}$.

	At the EOM-CCSDT/ANO1 level of theory used in the DVR, the \emph{cis} ground state lies above that of the \emph{trans} conformer by $\sim$2820 cm$^{-1}$, and its vibrational state manifold is consequently less dense at any given energy.  We find that all but the lowest-lying states contain at least a few percent \emph{trans} character, using the crude metric
	$2\int^{\pi}_{0}\int^{2\pi}_{\pi}\int^{\infty}_{0}|\psi|^{2}\ dR_{CC} d\phi_{1} d\phi_{2}$, but even these states are predicted to have non-negligible intensity in one of the four spectra discussed in Ref.~[\onlinecite{merer:054304}].
	These results are partially summarized in Table \ref{tab:cislevels}.  At higher energy it is frequently the case that one or both rovibrational symmetries interact so strongly with several \emph{trans} states that it is difficult to attach that zero-order assignment to any particular eigenstate.  Accordingly, it should be kept in mind that the mixing fractions are probably sensitively dependent on local resonances.  Spectral intensity is, as expected, generally correlated with \emph{trans} character, and, as noted earlier, excitation in $v_{3}$ and $v_{6}$ greatly enhances \emph{cis}-\emph{trans} mixing.  An in depth investigation of \emph{cis}-\emph{trans} interaction and intensity borrowing for a specific level is undertaken in the next section.
\begin{table}
\caption{Calculated \emph{cis} levels up to 5,000 cm$^{-1}$}
\vspace{8pt}
\begin{tabular*}{0.65\linewidth}{ c | c | c | c}
State\footnotemark[1] &\  $E - T_{0}^{trans}$&$E - T_{0}^{cis}$&\ \% \emph{trans}\\
\hline
$0^{0}$ & 2840.84 & 0 & $<$0.01\\
$6^{1}$ & 3429.19 & 588.35 &0.03\\
$3^{1}$ & 3630.4 & 789.56 &0.05\\
$6^{2}$ & 4015.4 & 1174.56 &1.61 \\
$3^{1}6^{1}$\footnotemark[2] & 4161.3 & 1320.46 &5.13\\
$2^{1}$ & 4330.45 & 1489.61 &0.17\\
$3^{2}$ & 4417.97 & 1577.13 &16.8\\
$6^{3}$ & 4576.07 & 1735.23 &25.9\\
$3^{1}6^{2}$ & 4664.0 & 1823.16 &45.5\\
$3^{2}6^{1}$ & 4883.01 & 2042.17 &34.7\\
$2^{1}6^{1}$ & 4919.7 & 2078.86 &2.01\\
\hline
\end{tabular*}
\footnotetext[1]{Values given are averaged between $oo$ and $ee$.}
\footnotetext[2]{This level will be discussed in detail in Sec.~\ref{sec:cis3161}.}
\label{tab:cislevels}
\end{table}
	
\subsection{Investigation of a Specific \emph{cis} $\leftrightarrow$ \emph{trans} Interaction}
\label{sec:cis3161}

	In order to determine the effects of \emph{cis} states on the \emph{trans} level structure and possible sources for their intensity, we would ideally like to compare the predictions of a model that neglects \emph{cis}-\emph{trans} mixing to the true spectrum (either calculated or experimental).
	In the absence of a global $\mathbf{H}^{\textrm{eff}}$ for the S$_{1}$ state, it is difficult to consider the level structure and spectral intensities in terms of a zero-order picture perturbed by the addition of an interaction.  However, we can use the DVR to approximate two non-interacting minima by performing two calculations wherein the wavefunctions are restricted to one geometry or the other.  The \emph{cis}-\emph{trans} ``interactions'' are then calculated by the trick of using the eigenstates of the full PES, $|\psi_{i}\rangle$, as a complete set to find the vibrational overlap integrals between the zero-order \emph{cis} and \emph{trans} wavefunctions
	\begin{equation}
		{}^{0}\langle \varphi_{m}^{cis}|\varphi_{n}^{trans}\rangle^{0}=\sum_{i} {}^{0}\langle \varphi_{m}^{cis}|\psi_{i}\rangle\langle\psi_{i}|\varphi_{n}^{trans}\rangle^{0}
	\end{equation}
and the mixing fractions follow from dividing by the zero-order energy differences.  (The ``true'' interaction matrix elements also include an electronic pre-factor, so that we obtain only relative coupling strengths.)  Zero-order spectra can also be computed between the $\tilde{X}$ state and the non-interacting sets of S$_{1}$ states, which allows for the tracking of intensity borrowing when the interaction is turned on.  It should also be noted that this method can aid in assigning highly mixed eigenstates, particularly by inspection of the quantities $\langle \psi_{i}|\varphi_{m}^{cis}\rangle^{0}$ and $\langle\psi_{i}|\varphi_{n}^{trans}\rangle^{0}$.

	The results of the above procedure as they apply to \emph{cis} $3^{1}6^{1}$, a state that lies in the region below \emph{trans} $3^{4}$, are discussed in the remainder of this section
	 \cite{footnote3}.  
	The vibrational $b_{2}$ symmetry of this state means that its $K=1$ level interacts with $K=1\ a_{g}$ levels, and that its $K=0,2$ levels interact with $K=0,2\ b_{u}$ levels.  The two rotational symmetries therefore require parallel but separate analyses, and so we will treat only $oo$, since it appears in the simpler one photon spectrum \cite{46175}.  
	
\begin{figure}
\includegraphics[width=\linewidth]{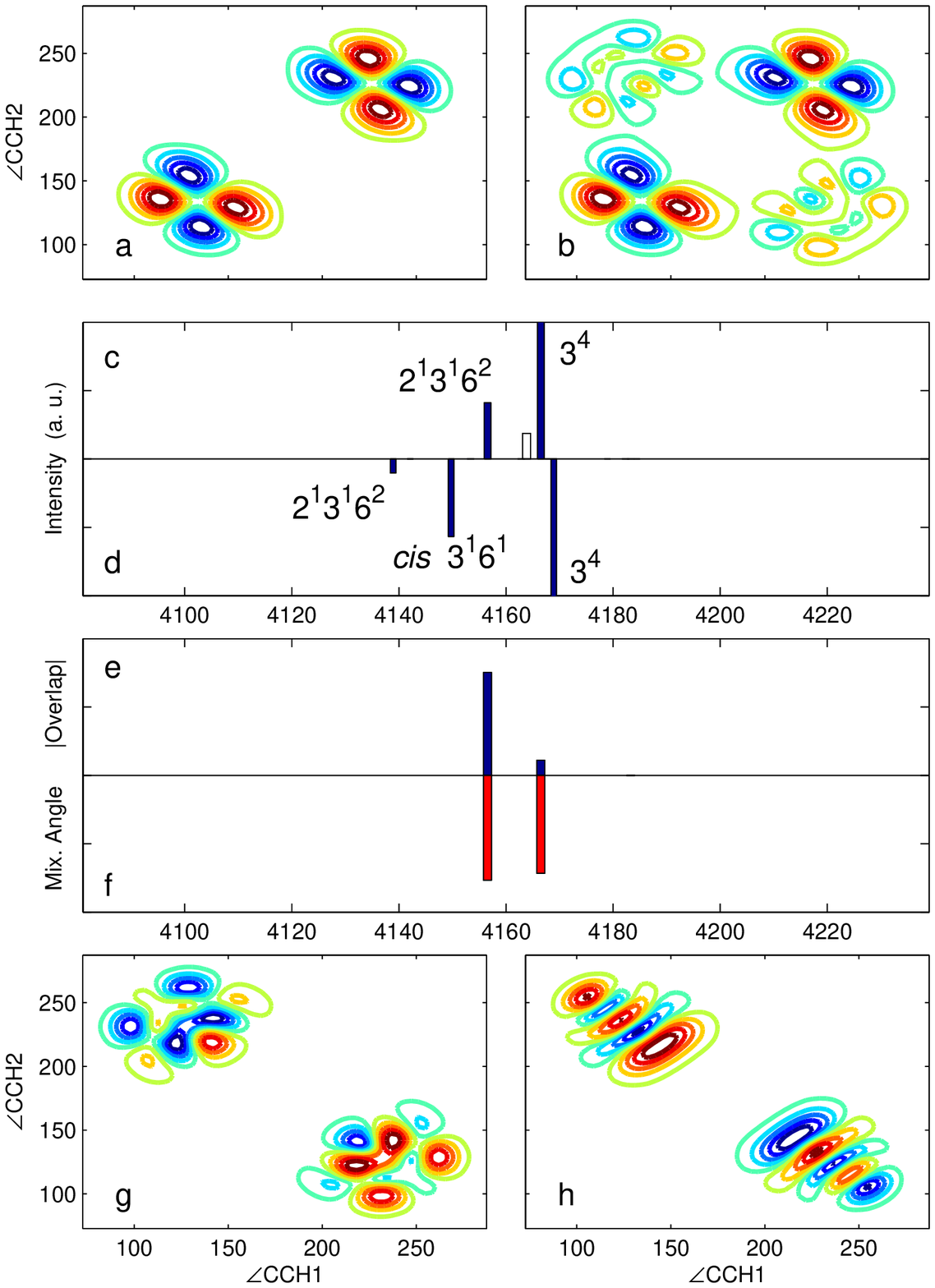}
\caption{
Plots depicting the analysis of \emph{cis}-\emph{trans} mixing and intensity borrowing involving  \emph{cis} $3^{1}6^{1}\ oo$.  (a)  Zero-order wavefunction of \emph{cis} $3^{1}6^{1}\ oo$.  (b)  Eigenstate with nominal assignment of \emph{cis} $3^{1}6^{1}\ oo$, showing significant \emph{trans} character. (c)  Stick spectrum of the zero-order states.  The peak of \emph{cis} $3^{1}6^{1}\ oo$ (white) has been multiplied by $10^{3}$. (d)  Stick spectrum of the eigenstates, in which intensity has been transferred to \emph{cis} $3^{1}6^{1}\ oo$ from the nearby \emph{trans} states.  (The intensity of \emph{trans} $3^{4}$ is off scale in both (c) and (d), and the energy offset between the two spectra is slightly arbitrary.) (e) Magnitudes of vibrational overlaps between the states shown in (c) and \emph{cis} $3^{1}6^{1}\ oo$.  Many other states (not shown) have as large or larger vibrational overlaps with {\it cis} $3^{1}6^{1}$.  (f) Magnitudes of the mixing angles  between the states shown in (c) and \emph{cis} $3^{1}6^{1}\ oo$.  These are the largest mixing angles by at least an order of magnitude, belonging to \emph{trans} $2^{1}3^{1}6^{2}$ and \emph{trans} $3^{4}$, whose zero-order wavefunctions are shown in (g) and (h), respectively.}
	\label{fig:cis3161}
\end{figure}

	The zero-order state of interest is trivially recognized by its nodal pattern (Fig.~\ref{fig:cis3161}a).  The eigenstate with this nominal assignment can then be identified (Fig.~\ref{fig:cis3161}b), even though the zero-order state is mixed into several eigenstates.  The spectra for the zero-order states and the eigenstates are plotted in Fig.~\ref{fig:cis3161}c-d for 100 cm$^{-1}$ above and below \emph{cis} $3^{1}6^{1}$.  These spectra differ in two diagnostically important ways.  First, the intensity has been redistributed such that the \emph{cis} state has increased its intensity from essentially nothing to greater than that of \emph{trans} $2^{1}3^{1}6^{2}$.  This tells us not only that there is \emph{cis}-\emph{trans} interaction, but also that the intensity of the \emph{cis} state derives entirely from this mixing and not from purely vibronic effects, such as the variation of the electronic transition moment with the nuclear coordinates (Fig.~\ref{fig:mu}).  Second, the energy level pattern has changed because the interacting states repel one another.  Interestingly, although ordinarily no strong resonance would mix \emph{trans} $2^{1}3^{1}6^{2}$ and \emph{trans} $3^{4}$, their interactions with a common \emph{cis} state cause them to move apart, an example of indirect mixing.  
	
	The explanation for these two phenomena is displayed in Fig.~\ref{fig:cis3161}e-f.  We see that \emph{trans} $2^{1}3^{1}6^{2}$ has a much larger overlap integral with \emph{cis} $3^{1}6^{1}$, as expected because of its excitation in $v_{6}$, but \emph{trans} $3^{4}$ is near resonant, and consequently the two states have approximately equal mixing angles with \emph{cis} $3^{1}6^{1}$.  This provides the \emph{cis} state with significant intensity, and in fact the calculated relative intensities for these three states agree quite well with those observed experimentally.  Finally, the vibrational overlaps between the two \emph{trans} states and the \emph{cis} state are in fact of opposite sign, which suggests the possibility of observable interference effects.  This three state interaction is currently being investigated experimentally, and early results are so far consistent with the calculation.


\subsection{Above-Barrier Dynamics}

\begin{table}
\caption{\emph{Ab initio} geometry and harmonic frequencies for $\tilde{A}\ ^{1}A''$ $^{12}$C$_{2}$H$_{2}$}
\vspace{8pt}
\begin{tabular*}{0.4\linewidth}{ c | c }
$\omega_{1}$ &\ 3405.17\\
$\omega_{2}$ &\ 2745.18\\
$\omega_{3}$ &\ 1470.48\\
$\omega_{4}$ &\ 890.01\\
$\omega_{5}$ &\ 886.19$i$\\
$\omega_{6}$ &\ 766.18\\
\hline
$\angle$CCH$_{1}$ \ &  119.80\\
$\angle$CCH$_{2}$ \ &  178.70\\
$R_{\textrm{CH}_{1}}$ &  1.1147\\
$R_{\textrm{CH}_{2}}$ &  1.0680\\
$R_{\textrm{CC}}$ &  1.3548\\
$T_{e}-T_{e}^{trans}$ & 4979\\
\end{tabular*}
\footnotetext[0]{Frequencies in cm$^{-1}$, angles in degrees, and bond lengths in \AA.  The transition state geometry is slightly \emph{cis}-bent.}
\label{tab:TSabinitio}
\end{table}

\begin{figure}
\includegraphics[width=\linewidth]{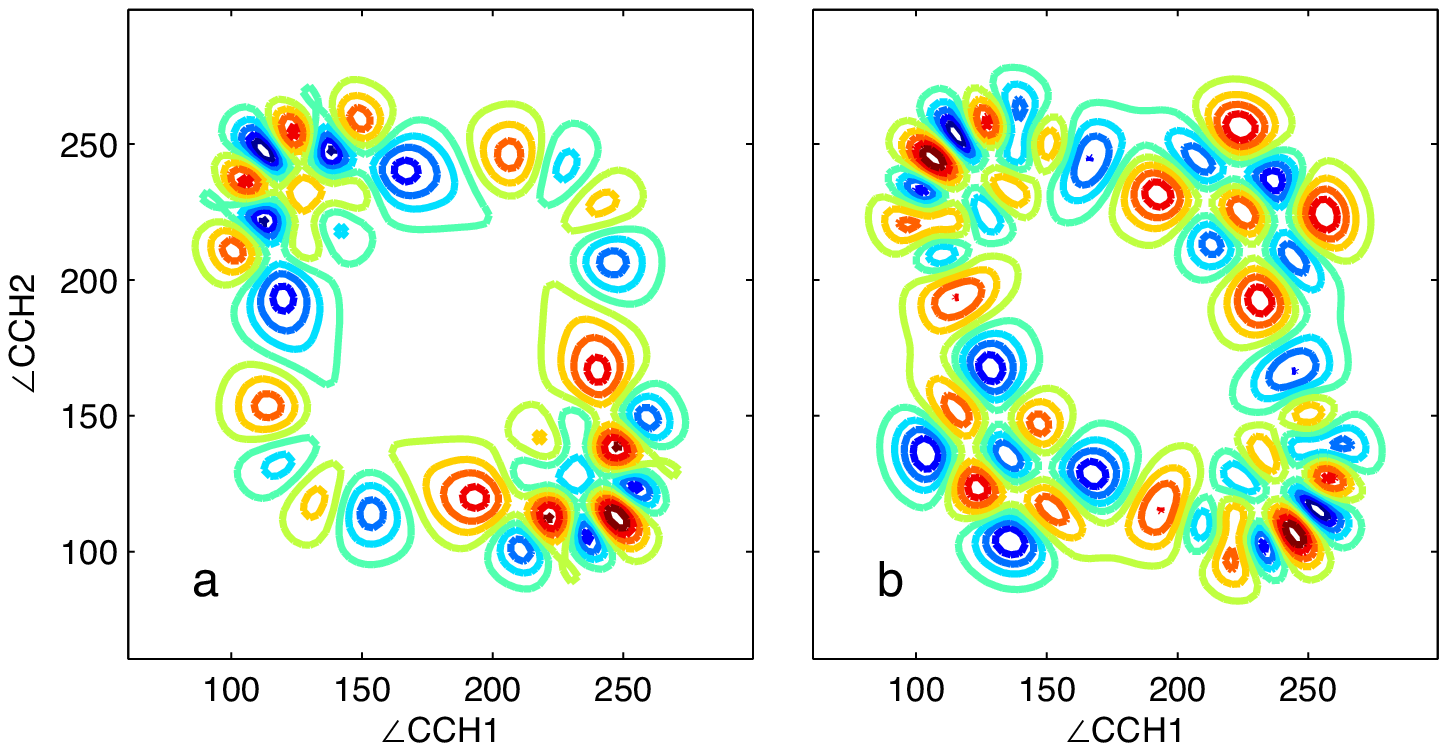}
\caption{
Wavefunctions of delocalized states above the isomerization barrier.}
	\label{fig:above}
\end{figure}

        A full dimensional transition state search finds the \emph{cis}-\emph{trans} barrier height to be 4979 cm$^{-1}$ (Table \ref{tab:TSabinitio}).  In the reduced dimension PES used in the DVR, the barrier height is effectively 5145 cm$^{-1}$.
         \cite{footnote4}  Above this energy, delocalized states begin to appear \cite{2001PCCP....3.5393M} (Fig.~\ref{fig:above}), but other states unrelated to the isomerization coordinate remain unaffected.  Although any conclusions about the above-barrier dynamics would be premature, a cursory analysis of the delocalized wavefunctions shows that they correspond to out of phase combinations of local benders.  The elliptical shapes (``ring modes'') are therefore similar to Lissajous figures for two equal amplitude oscillations with a phase difference of $\pm\pi/4$ or $\pm3\pi/4$, unlike the patterns along diagonal or cardinal axes that accompany normal or local mode behavior, respectively (Fig.~\ref{fig:normallocal}).  The two differently inclined ring modes are presumably the above-barrier counterparts of \emph{cis} and \emph{trans} bending.


	Some states that have amplitude in both wells do not exhibit this ring mode behavior; in our results such states always have nodal patterns in each well that are clearly assignable, indicating that these states are less affected by the isomerization despite their delocalization.
	
	The shapes of the wavefunctions both above and below the barrier to isomerization provide some clues about the important resonances in the bending Hamiltonian.  Since we do not observe the cross shaped wavefunctions indicative of local modes, we can infer that the $K_{3366}$ Darling--Dennison resonance does not dominate the dynamics.  This is in line with the frequency ratio of the two modes being not $2:2$ but rather $\sim 1.3:2$, lying in between the usual ratios for strong Darling--Dennison and Fermi resonance.  We find \cite{steevesthesis} that both types of resonance are necessary in an $\mathbf{H}^{\textrm{eff}}$ to reproduce the unusual nodal patterns of the below-barrier \emph{trans} well DVR wavefunctions.  The presence of both strong Fermi and Darling--Dennison resonance leads to the destruction of any polyad structure in the $\tilde{A}$ state.  For fixed values of ($v_{1}$, $v_{2}$, $v_{5}$), all states of a given symmetry can interact via known anharmonic interactions, so the Hamiltonian does
not readily block diagonalize according to conserved polyad quantum numbers.  Research into this phenomenon and its relation to low barrier isomerization is ongoing.

\section{Conclusion}
	We set out in this paper to investigate the spectroscopic consequences of low barrier \emph{cis}-\emph{trans} isomerization in S$_{1}$ acetylene by calculating the vibrational eigenstates of a high accuracy PES using a reduced dimension DVR method.  The calculation reproduces some difficult aspects of the \emph{trans} conformer level structure, which are ultimately due to the isomerization.  Another consequence of the isomerization process is that nominally forbidden transitions to the \emph{cis} conformer appear.  These transitions occur in our calculation near the observed energies of the ``extra'' levels in the $\tilde{A}\leftarrow\tilde{X}$ spectrum of acetylene.  We find that our reduced dimension DVR calculation agrees with the explanation of these ``extra'' levels as belonging to the \emph{cis} conformer, and that we are able to explain the intensities of these levels by using the calculated results to investigate \emph{cis}-\emph{trans} mixings.  Although we believe that the reduced dimension DVR method demonstrated here is a powerful approach for studying isomerizing systems, there are interesting aspects of the S$_{1}$ C$_{2}$H$_{2}$ spectrum that it does not address, including strong vibration-rotation interactions and possible multiple pathways to isomerization.  We intend to develop a full dimensional treatment for S$_{1}$ acetylene to study these effects and enable comparison with the complete set of experimental observations.

\acknowledgments
J. Baraban would like to thank D.~Tannor, T.~Van Voorhis and A.~Merer for helpful discussions, and is grateful for support by an NSF Graduate Research Fellowship.  This work was supported at MIT by DOE Grant No. DE-FG0287ER13671.

\appendix
\section{Computation of S$_{1}\leftrightarrow$ S$_{0}$ Spectra}
In order to calculate $\tilde{A}\leftrightarrow\tilde{X}$ spectra, two additional quantities are required beyond the $\tilde{A}$ state vibrational wavefunctions: the electronic transition moment, $\mu$, and the $\tilde{X}$ state vibrational wavefunctions.  We calculate the transition moment \emph{ab initio} over our coordinate grid at the EOM-CCSD/ANO1 level of theory, again using the CFOUR program system \cite{cfour}. The $\tilde{X}$ state vibrational wavefunctions can be obtained merely by replacing the $\tilde{A}$ state PES with that of the $\tilde{X}$ state and repeating the DVR calculation.  This portability of the method is one of its powerful features.  Although the results thus obtained for the $\tilde{X}$ state are not worth examining in depth, given the numerous \emph{ab initio} treatments in the literature, we would like to note that the qualitatively important features related to large amplitude motions are simulated correctly by our DVR method.  Chief among these is the emergence of the ``local-bender'' states \cite{jacobson:845,sibert:937} from the normal modes $v_{4}''$ and $v_{5}''$, as shown in Fig.~\ref{fig:normallocal}.


\begin{figure}
\includegraphics[width=\linewidth]{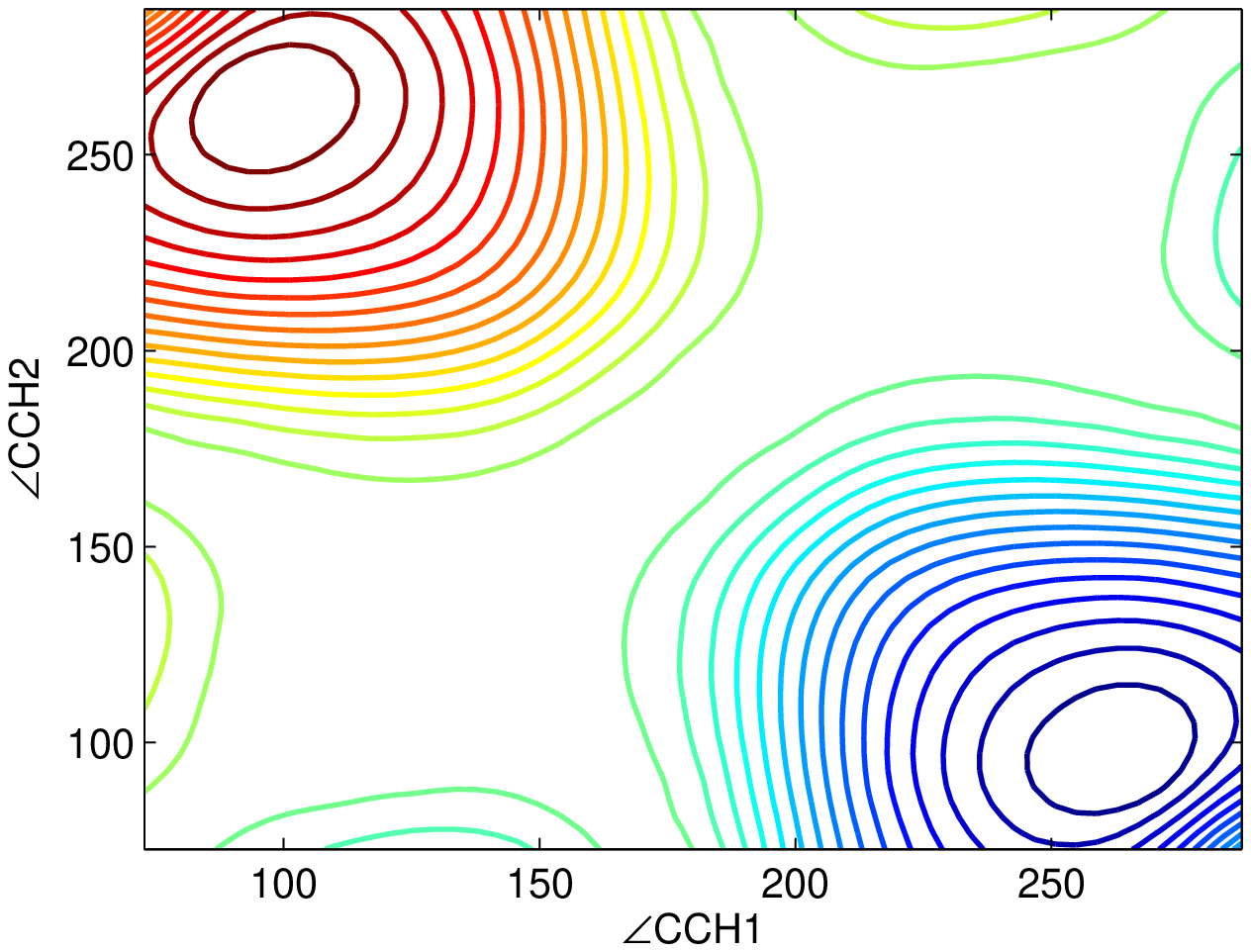}
\caption{
The transition moment $\mu$ as a function of the bending angles.  Note that $\mu=0$ at all points of C$_{2v}$ symmetry, including linearity.  In order for the overall symmetry of $\mu$ to be $Ss-$, the quantity used to calculate $\tilde{A}\leftrightarrow\tilde{X}$ spectra must have $Sa-$ symmetry in the DVR coordinate space.  Briefly, this requirement arises because the $\tilde{X}$ state DVR wavefunctions are vibrational instead of rovibrational.  Adding the rotational symmetry factor necessary to ensure that transitions occur between the same nuclear spin symmetry species includes an extra $Sa-$ in $\mu$.}
	\label{fig:mu}
\end{figure}


\begin{figure}
\includegraphics[width=\linewidth]{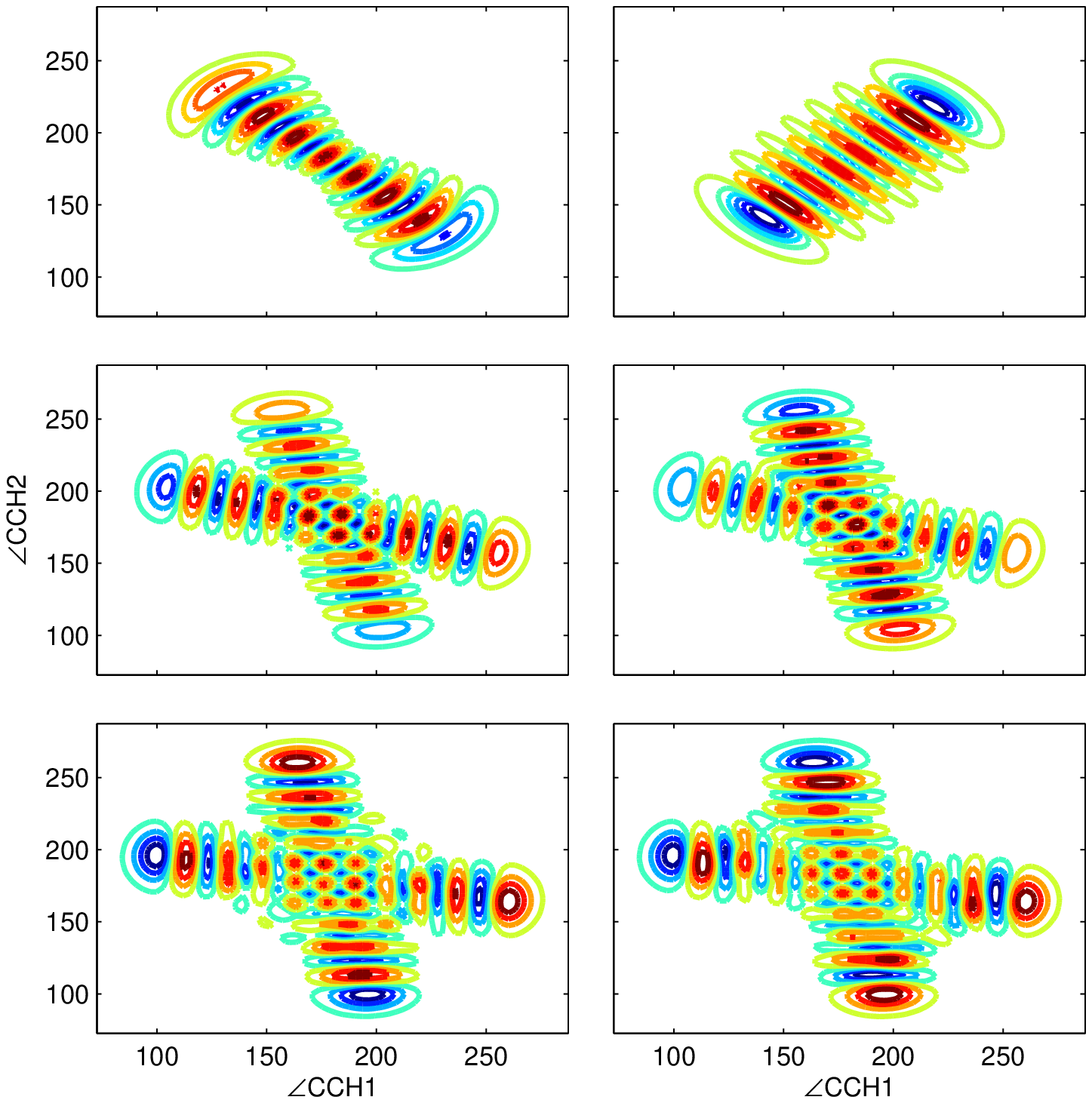}
\caption{
From top to bottom, pairs of wavefunctions depicting the evolution from the normal mode limit to the local mode limit in S$_{0}$ acetylene from approximately $8,000$ to $12,000$ cm$^{-1}$ of vibrational excitation.}
	\label{fig:normallocal}
\end{figure}

With these two results in hand, we calculate for each pair of eigenstates the spectral intensity $I_{mn} = |\langle m|\mu|n\rangle|^{2}$, where $m$ and $n$ belong to different electronic states.  In order to calculate the product it is necessary to interpolate the transition moment and wavefunctions to a common set of grid points.  This process yields all possible upward spectra, including the four necessary to sample experimentally the full rovibrational structure of the $\tilde{A}$ state \cite{merer:054304}, and also all possible downward spectra (DF/SEP).  This wealth of information was of enormous value when analyzing the calculated results, and we expect it to guide future experiments.

\bibliography{DVRfoot}

\begin{thebibliography}{53}%
\makeatletter
\providecommand \@ifxundefined [1]{%
 \@ifx{#1\undefined}
}%
\providecommand \@ifnum [1]{%
 \ifnum #1\expandafter \@firstoftwo
 \else \expandafter \@secondoftwo
 \fi
}%
\providecommand \@ifx [1]{%
 \ifx #1\expandafter \@firstoftwo
 \else \expandafter \@secondoftwo
 \fi
}%
\providecommand \natexlab [1]{#1}%
\providecommand \enquote  [1]{``#1''}%
\providecommand \bibnamefont  [1]{#1}%
\providecommand \bibfnamefont [1]{#1}%
\providecommand \citenamefont [1]{#1}%
\providecommand \href@noop [0]{\@secondoftwo}%
\providecommand \href [0]{\begingroup \@sanitize@url \@href}%
\providecommand \@href[1]{\@@startlink{#1}\@@href}%
\providecommand \@@href[1]{\endgroup#1\@@endlink}%
\providecommand \@sanitize@url [0]{\catcode `\\12\catcode `\$12\catcode
  `\&12\catcode `\#12\catcode `\^12\catcode `\_12\catcode `\%12\relax}%
\providecommand \@@startlink[1]{}%
\providecommand \@@endlink[0]{}%
\providecommand \url  [0]{\begingroup\@sanitize@url \@url }%
\providecommand \@url [1]{\endgroup\@href {#1}{\urlprefix }}%
\providecommand \urlprefix  [0]{URL }%
\providecommand \Eprint [0]{\href }%
\providecommand \doibase [0]{http://dx.doi.org/}%
\providecommand \selectlanguage [0]{\@gobble}%
\providecommand \bibinfo  [0]{\@secondoftwo}%
\providecommand \bibfield  [0]{\@secondoftwo}%
\providecommand \translation [1]{[#1]}%
\providecommand \BibitemOpen [0]{}%
\providecommand \bibitemStop [0]{}%
\providecommand \bibitemNoStop [0]{.\EOS\space}%
\providecommand \EOS [0]{\spacefactor3000\relax}%
\providecommand \BibitemShut  [1]{\csname bibitem#1\endcsname}%
\let\auto@bib@innerbib\@empty
\bibitem [{\citenamefont {King}\ and\ \citenamefont
  {Ingold}(1952)}]{king_bent_1952}%
  \BibitemOpen
  \bibfield  {author} {\bibinfo {author} {\bibfnamefont {G.~W.}\ \bibnamefont
  {King}}\ and\ \bibinfo {author} {\bibfnamefont {C.~K.}\ \bibnamefont
  {Ingold}},\ }\href {\doibase 10.1038/1691101b0} {\bibfield  {journal}
  {\bibinfo  {journal} {Nature}\ }\textbf {\bibinfo {volume} {169}},\ \bibinfo
  {pages} {1101} (\bibinfo {year} {1952})}\BibitemShut {NoStop}%
\bibitem [{\citenamefont {Ingold}\ and\ \citenamefont
  {King}(1953)}]{ingold1953esa}%
  \BibitemOpen
  \bibfield  {author} {\bibinfo {author} {\bibfnamefont {C.~K.}\ \bibnamefont
  {Ingold}}\ and\ \bibinfo {author} {\bibfnamefont {G.~W.}\ \bibnamefont
  {King}},\ }\href@noop {} {\bibfield  {journal} {\bibinfo  {journal} {J. Chem.
  Soc.}\ ,\ \bibinfo {pages} {2702}} (\bibinfo {year} {1953})}\BibitemShut
  {NoStop}%
\bibitem [{\citenamefont {Innes}(1954)}]{innes:863}%
  \BibitemOpen
  \bibfield  {author} {\bibinfo {author} {\bibfnamefont {K.~K.}\ \bibnamefont
  {Innes}},\ }\href {\doibase 10.1063/1.1740204} {\bibfield  {journal}
  {\bibinfo  {journal} {The Journal of Chemical Physics}\ }\textbf {\bibinfo
  {volume} {22}},\ \bibinfo {pages} {863} (\bibinfo {year} {1954})}\BibitemShut
  {NoStop}%
\bibitem [{\citenamefont {Hougen}\ and\ \citenamefont
  {Watson}(1965)}]{hougen1965anomalous}%
  \BibitemOpen
  \bibfield  {author} {\bibinfo {author} {\bibfnamefont {J.}~\bibnamefont
  {Hougen}}\ and\ \bibinfo {author} {\bibfnamefont {J.}~\bibnamefont
  {Watson}},\ }\href@noop {} {\bibfield  {journal} {\bibinfo  {journal}
  {Canadian Journal of Physics}\ }\textbf {\bibinfo {volume} {43}},\ \bibinfo
  {pages} {298} (\bibinfo {year} {1965})}\BibitemShut {NoStop}%
\bibitem [{\citenamefont {Abramson}\ \emph {et~al.}(1982)\citenamefont
  {Abramson}, \citenamefont {Kittrell}, \citenamefont {Kinsey},\ and\
  \citenamefont {Field}}]{abramson:2293}%
  \BibitemOpen
  \bibfield  {author} {\bibinfo {author} {\bibfnamefont {E.}~\bibnamefont
  {Abramson}}, \bibinfo {author} {\bibfnamefont {C.}~\bibnamefont {Kittrell}},
  \bibinfo {author} {\bibfnamefont {J.~L.}\ \bibnamefont {Kinsey}}, \ and\
  \bibinfo {author} {\bibfnamefont {R.~W.}\ \bibnamefont {Field}},\ }\href
  {\doibase 10.1063/1.443301} {\bibfield  {journal} {\bibinfo  {journal} {The
  Journal of Chemical Physics}\ }\textbf {\bibinfo {volume} {76}},\ \bibinfo
  {pages} {2293} (\bibinfo {year} {1982})}\BibitemShut {NoStop}%
\bibitem [{\citenamefont {Dupr{\'e}}\ \emph {et~al.}(1991)\citenamefont
  {Dupr{\'e}}, \citenamefont {Jost}, \citenamefont {Lombardi}, \citenamefont
  {Green}, \citenamefont {Abramson},\ and\ \citenamefont
  {Field}}]{Dupre1991293}%
  \BibitemOpen
  \bibfield  {author} {\bibinfo {author} {\bibfnamefont {P.}~\bibnamefont
  {Dupr{\'e}}}, \bibinfo {author} {\bibfnamefont {R.}~\bibnamefont {Jost}},
  \bibinfo {author} {\bibfnamefont {M.}~\bibnamefont {Lombardi}}, \bibinfo
  {author} {\bibfnamefont {P.}~\bibnamefont {Green}}, \bibinfo {author}
  {\bibfnamefont {E.}~\bibnamefont {Abramson}}, \ and\ \bibinfo {author}
  {\bibfnamefont {R.}~\bibnamefont {Field}},\ }\href {\doibase DOI:
  10.1016/0301-0104(91)85006-3} {\bibfield  {journal} {\bibinfo  {journal}
  {Chemical Physics}\ }\textbf {\bibinfo {volume} {152}},\ \bibinfo {pages}
  {293 } (\bibinfo {year} {1991})}\BibitemShut {NoStop}%
\bibitem [{\citenamefont {Drabbels}, \citenamefont {Heinze},\ and\
  \citenamefont {Meerts}(1994)}]{drabbels:165}%
  \BibitemOpen
  \bibfield  {author} {\bibinfo {author} {\bibfnamefont {M.}~\bibnamefont
  {Drabbels}}, \bibinfo {author} {\bibfnamefont {J.}~\bibnamefont {Heinze}}, \
  and\ \bibinfo {author} {\bibfnamefont {W.~L.}\ \bibnamefont {Meerts}},\
  }\href {\doibase 10.1063/1.466988} {\bibfield  {journal} {\bibinfo  {journal}
  {The Journal of Chemical Physics}\ }\textbf {\bibinfo {volume} {100}},\
  \bibinfo {pages} {165} (\bibinfo {year} {1994})}\BibitemShut {NoStop}%
\bibitem [{\citenamefont {Utz}\ \emph {et~al.}(1993)\citenamefont {Utz},
  \citenamefont {Tobiason}, \citenamefont {M.}, \citenamefont {Sanders},\ and\
  \citenamefont {Crim}}]{utz:2742}%
  \BibitemOpen
  \bibfield  {author} {\bibinfo {author} {\bibfnamefont {A.~L.}\ \bibnamefont
  {Utz}}, \bibinfo {author} {\bibfnamefont {J.~D.}\ \bibnamefont {Tobiason}},
  \bibinfo {author} {\bibfnamefont {E.~C.}\ \bibnamefont {M.}}, \bibinfo
  {author} {\bibfnamefont {L.~J.}\ \bibnamefont {Sanders}}, \ and\ \bibinfo
  {author} {\bibfnamefont {F.~F.}\ \bibnamefont {Crim}},\ }\href {\doibase
  10.1063/1.464156} {\bibfield  {journal} {\bibinfo  {journal} {The Journal of
  Chemical Physics}\ }\textbf {\bibinfo {volume} {98}},\ \bibinfo {pages}
  {2742} (\bibinfo {year} {1993})}\BibitemShut {NoStop}%
\bibitem [{\citenamefont {Merer}\ \emph {et~al.}(2008)\citenamefont {Merer},
  \citenamefont {Yamakita}, \citenamefont {Tsuchiya}, \citenamefont {Steeves},
  \citenamefont {Bechtel},\ and\ \citenamefont {Field}}]{merer:054304}%
  \BibitemOpen
  \bibfield  {author} {\bibinfo {author} {\bibfnamefont {A.~J.}\ \bibnamefont
  {Merer}}, \bibinfo {author} {\bibfnamefont {N.}~\bibnamefont {Yamakita}},
  \bibinfo {author} {\bibfnamefont {S.}~\bibnamefont {Tsuchiya}}, \bibinfo
  {author} {\bibfnamefont {A.~H.}\ \bibnamefont {Steeves}}, \bibinfo {author}
  {\bibfnamefont {H.~A.}\ \bibnamefont {Bechtel}}, \ and\ \bibinfo {author}
  {\bibfnamefont {R.~W.}\ \bibnamefont {Field}},\ }\href {\doibase
  10.1063/1.2939246} {\bibfield  {journal} {\bibinfo  {journal} {The Journal of
  Chemical Physics}\ }\textbf {\bibinfo {volume} {129}},\ \bibinfo {eid}
  {054304} (\bibinfo {year} {2008})}\BibitemShut {NoStop}%
\bibitem [{\citenamefont {Steeves}\ \emph {et~al.}(2009)\citenamefont
  {Steeves}, \citenamefont {Bechtel}, \citenamefont {Merer}, \citenamefont
  {Yamakita}, \citenamefont {Tsuchiya},\ and\ \citenamefont
  {Field}}]{steeves2009stretchbend}%
  \BibitemOpen
  \bibfield  {author} {\bibinfo {author} {\bibfnamefont {A.}~\bibnamefont
  {Steeves}}, \bibinfo {author} {\bibfnamefont {H.}~\bibnamefont {Bechtel}},
  \bibinfo {author} {\bibfnamefont {A.}~\bibnamefont {Merer}}, \bibinfo
  {author} {\bibfnamefont {N.}~\bibnamefont {Yamakita}}, \bibinfo {author}
  {\bibfnamefont {S.}~\bibnamefont {Tsuchiya}}, \ and\ \bibinfo {author}
  {\bibfnamefont {R.}~\bibnamefont {Field}},\ }\href@noop {} {\bibfield
  {journal} {\bibinfo  {journal} {Journal of Molecular Spectroscopy}\ }\textbf
  {\bibinfo {volume} {256}},\ \bibinfo {pages} {256} (\bibinfo {year}
  {2009})}\BibitemShut {NoStop}%
\bibitem [{\citenamefont {Merer}\ \emph {et~al.}(2010)\citenamefont {Merer},
  \citenamefont {Steeves}, \citenamefont {Baraban}, \citenamefont {Bechtel},\
  and\ \citenamefont {Field}}]{46175}%
  \BibitemOpen
  \bibfield  {author} {\bibinfo {author} {\bibfnamefont {A.~J.}\ \bibnamefont
  {Merer}}, \bibinfo {author} {\bibfnamefont {A.~H.}\ \bibnamefont {Steeves}},
  \bibinfo {author} {\bibfnamefont {J.~H.}\ \bibnamefont {Baraban}}, \bibinfo
  {author} {\bibfnamefont {H.~A.}\ \bibnamefont {Bechtel}}, \ and\ \bibinfo
  {author} {\bibfnamefont {R.~W.}\ \bibnamefont {Field}},\ }\href@noop {}
  {\bibfield  {journal} {\bibinfo  {journal} {in preparation}\ } (\bibinfo
  {year} {2010})}\BibitemShut {NoStop}%
\bibitem [{\citenamefont {Demoulin}(1975)}]{Demoulin1975329}%
  \BibitemOpen
  \bibfield  {author} {\bibinfo {author} {\bibfnamefont {D.}~\bibnamefont
  {Demoulin}},\ }\href {\doibase DOI: 10.1016/0301-0104(75)80013-9} {\bibfield
  {journal} {\bibinfo  {journal} {Chemical Physics}\ }\textbf {\bibinfo
  {volume} {11}},\ \bibinfo {pages} {329 } (\bibinfo {year}
  {1975})}\BibitemShut {NoStop}%
\bibitem [{\citenamefont {Stanton}, \citenamefont {Huang},\ and\ \citenamefont
  {Szalay}(1994)}]{stanton:356}%
  \BibitemOpen
  \bibfield  {author} {\bibinfo {author} {\bibfnamefont {J.~F.}\ \bibnamefont
  {Stanton}}, \bibinfo {author} {\bibfnamefont {C.-M.}\ \bibnamefont {Huang}},
  \ and\ \bibinfo {author} {\bibfnamefont {P.~G.}\ \bibnamefont {Szalay}},\
  }\href {\doibase 10.1063/1.468142} {\bibfield  {journal} {\bibinfo  {journal}
  {The Journal of Chemical Physics}\ }\textbf {\bibinfo {volume} {101}},\
  \bibinfo {pages} {356} (\bibinfo {year} {1994})}\BibitemShut {NoStop}%
\bibitem [{\citenamefont {Ventura}, \citenamefont {Dallos},\ and\ \citenamefont
  {Lischka}(2003)}]{ventura2003ves}%
  \BibitemOpen
  \bibfield  {author} {\bibinfo {author} {\bibfnamefont {E.}~\bibnamefont
  {Ventura}}, \bibinfo {author} {\bibfnamefont {M.}~\bibnamefont {Dallos}}, \
  and\ \bibinfo {author} {\bibfnamefont {H.}~\bibnamefont {Lischka}},\
  }\href@noop {} {\bibfield  {journal} {\bibinfo  {journal} {The Journal of
  Chemical Physics}\ }\textbf {\bibinfo {volume} {118}},\ \bibinfo {pages}
  {1702} (\bibinfo {year} {2003})}\BibitemShut {NoStop}%
\bibitem [{\citenamefont {{Malsch}}\ \emph {et~al.}(2001)\citenamefont
  {{Malsch}}, \citenamefont {{Hohlneicher}}, \citenamefont {{Schork}},\ and\
  \citenamefont {{K{\"o}ppel}}}]{2001PCCP....3.5393M}%
  \BibitemOpen
  \bibfield  {author} {\bibinfo {author} {\bibfnamefont {K.}~\bibnamefont
  {{Malsch}}}, \bibinfo {author} {\bibfnamefont {G.}~\bibnamefont
  {{Hohlneicher}}}, \bibinfo {author} {\bibfnamefont {R.}~\bibnamefont
  {{Schork}}}, \ and\ \bibinfo {author} {\bibfnamefont {H.}~\bibnamefont
  {{K{\"o}ppel}}},\ }\href {\doibase 10.1039/b106391j} {\bibfield  {journal}
  {\bibinfo  {journal} {Physical Chemistry Chemical Physics (Incorporating
  Faraday Transactions)}\ }\textbf {\bibinfo {volume} {3}},\ \bibinfo {pages}
  {5393} (\bibinfo {year} {2001})}\BibitemShut {NoStop}%
\bibitem [{\citenamefont {Schubert}, \citenamefont {K{\"o}ppel},\ and\
  \citenamefont {Lischka}(2005)}]{Schubert:2005fk}%
  \BibitemOpen
  \bibfield  {author} {\bibinfo {author} {\bibfnamefont {B.}~\bibnamefont
  {Schubert}}, \bibinfo {author} {\bibfnamefont {H.}~\bibnamefont
  {K{\"o}ppel}}, \ and\ \bibinfo {author} {\bibfnamefont {H.}~\bibnamefont
  {Lischka}},\ }\href {\doibase 10.1063/1.1890865} {\bibfield  {journal}
  {\bibinfo  {journal} {J Chem Phys}\ }\textbf {\bibinfo {volume} {122}},\
  \bibinfo {pages} {184312} (\bibinfo {year} {2005})}\BibitemShut {NoStop}%
\bibitem [{\citenamefont {Light}\ and\ \citenamefont
  {Carrington~Jr}(2000)}]{light2000dvr}%
  \BibitemOpen
  \bibfield  {author} {\bibinfo {author} {\bibfnamefont {J.}~\bibnamefont
  {Light}}\ and\ \bibinfo {author} {\bibfnamefont {T.}~\bibnamefont
  {Carrington~Jr}},\ }\href@noop {} {\bibfield  {journal} {\bibinfo  {journal}
  {Advances in Chemical Physics}\ }\textbf {\bibinfo {volume} {114}},\ \bibinfo
  {pages} {263} (\bibinfo {year} {2000})}\BibitemShut {NoStop}%
\bibitem [{\citenamefont {Baci\'{c}}\ and\ \citenamefont
  {Light}(1986)}]{bacic:4594}%
  \BibitemOpen
  \bibfield  {author} {\bibinfo {author} {\bibfnamefont {Z.}~\bibnamefont
  {Baci\'{c}}}\ and\ \bibinfo {author} {\bibfnamefont {J.~C.}\ \bibnamefont
  {Light}},\ }\href {\doibase 10.1063/1.451824} {\bibfield  {journal} {\bibinfo
   {journal} {The Journal of Chemical Physics}\ }\textbf {\bibinfo {volume}
  {85}},\ \bibinfo {pages} {4594} (\bibinfo {year} {1986})}\BibitemShut
  {NoStop}%
\bibitem [{\citenamefont {{Kucharski}}\ \emph {et~al.}(2001)\citenamefont
  {{Kucharski}}, \citenamefont {{W{\l}och}}, \citenamefont {{Musia{\l}}},\ and\
  \citenamefont {{Bartlett}}}]{2001JChPh.115.8263K}%
  \BibitemOpen
  \bibfield  {author} {\bibinfo {author} {\bibfnamefont {S.~A.}\ \bibnamefont
  {{Kucharski}}}, \bibinfo {author} {\bibfnamefont {M.}~\bibnamefont
  {{W{\l}och}}}, \bibinfo {author} {\bibfnamefont {M.}~\bibnamefont
  {{Musia{\l}}}}, \ and\ \bibinfo {author} {\bibfnamefont {R.~J.}\ \bibnamefont
  {{Bartlett}}},\ }\href {\doibase 10.1063/1.1416173} {\bibfield  {journal}
  {\bibinfo  {journal} {Journal of Chemical Physics}\ }\textbf {\bibinfo
  {volume} {115}},\ \bibinfo {pages} {8263} (\bibinfo {year}
  {2001})}\BibitemShut {NoStop}%
\bibitem [{\citenamefont {Bentley}\ \emph {et~al.}(1992)\citenamefont
  {Bentley}, \citenamefont {Wyatt}, \citenamefont {Menou},\ and\ \citenamefont
  {Leforestier}}]{bentley:4255}%
  \BibitemOpen
  \bibfield  {author} {\bibinfo {author} {\bibfnamefont {J.~A.}\ \bibnamefont
  {Bentley}}, \bibinfo {author} {\bibfnamefont {R.~E.}\ \bibnamefont {Wyatt}},
  \bibinfo {author} {\bibfnamefont {M.}~\bibnamefont {Menou}}, \ and\ \bibinfo
  {author} {\bibfnamefont {C.}~\bibnamefont {Leforestier}},\ }\href {\doibase
  10.1063/1.463927} {\bibfield  {journal} {\bibinfo  {journal} {The Journal of
  Chemical Physics}\ }\textbf {\bibinfo {volume} {97}},\ \bibinfo {pages}
  {4255} (\bibinfo {year} {1992})}\BibitemShut {NoStop}%
\bibitem [{\citenamefont {III}\ and\ \citenamefont
  {Mayrhofer}(1993)}]{sibert:937}%
  \BibitemOpen
  \bibfield  {author} {\bibinfo {author} {\bibfnamefont {E.~L.~S.}\
  \bibnamefont {III}}\ and\ \bibinfo {author} {\bibfnamefont {R.~C.}\
  \bibnamefont {Mayrhofer}},\ }\href {\doibase 10.1063/1.465358} {\bibfield
  {journal} {\bibinfo  {journal} {The Journal of Chemical Physics}\ }\textbf
  {\bibinfo {volume} {99}},\ \bibinfo {pages} {937} (\bibinfo {year}
  {1993})}\BibitemShut {NoStop}%
\bibitem [{\citenamefont {Bunker}\ and\ \citenamefont
  {Jensen}(2006)}]{bunker2006msa}%
  \BibitemOpen
  \bibfield  {author} {\bibinfo {author} {\bibfnamefont {P.}~\bibnamefont
  {Bunker}}\ and\ \bibinfo {author} {\bibfnamefont {P.}~\bibnamefont
  {Jensen}},\ }\href@noop {} {\emph {\bibinfo {title} {{Molecular Symmetry and
  Spectroscopy}}}}\ (\bibinfo  {publisher} {NRC Research Press},\ \bibinfo
  {year} {2006})\BibitemShut {NoStop}%
\bibitem [{\citenamefont {Prosmiti}\ and\ \citenamefont
  {Farantos}(1995)}]{prosmiti:3299}%
  \BibitemOpen
  \bibfield  {author} {\bibinfo {author} {\bibfnamefont {R.}~\bibnamefont
  {Prosmiti}}\ and\ \bibinfo {author} {\bibfnamefont {S.~C.}\ \bibnamefont
  {Farantos}},\ }\href {\doibase 10.1063/1.470264} {\bibfield  {journal}
  {\bibinfo  {journal} {The Journal of Chemical Physics}\ }\textbf {\bibinfo
  {volume} {103}},\ \bibinfo {pages} {3299} (\bibinfo {year}
  {1995})}\BibitemShut {NoStop}%
\bibitem [{\citenamefont {Bramley}, \citenamefont {Green},\ and\ \citenamefont
  {Handy}(1991)}]{bramley1991vrc}%
  \BibitemOpen
  \bibfield  {author} {\bibinfo {author} {\bibfnamefont {M.}~\bibnamefont
  {Bramley}}, \bibinfo {author} {\bibfnamefont {W.}~\bibnamefont {Green}}, \
  and\ \bibinfo {author} {\bibfnamefont {N.}~\bibnamefont {Handy}},\
  }\href@noop {} {\bibfield  {journal} {\bibinfo  {journal} {Molecular
  Physics}\ }\textbf {\bibinfo {volume} {73}},\ \bibinfo {pages} {1183}
  (\bibinfo {year} {1991})}\BibitemShut {NoStop}%
\bibitem [{\citenamefont {Wilson}, \citenamefont {Decius},\ and\ \citenamefont
  {Cross}(1980)}]{wilson1980mvt}%
  \BibitemOpen
  \bibfield  {author} {\bibinfo {author} {\bibfnamefont {E.~B.}\ \bibnamefont
  {Wilson}}, \bibinfo {author} {\bibfnamefont {J.~C.}\ \bibnamefont {Decius}},
  \ and\ \bibinfo {author} {\bibfnamefont {P.~C.}\ \bibnamefont {Cross}},\
  }\href@noop {} {\emph {\bibinfo {title} {{Molecular Vibrations: The Theory of
  Infrared and Raman Vibrational Spectra}}}}\ (\bibinfo  {publisher} {Courier
  Dover Publications},\ \bibinfo {year} {1980})\BibitemShut {NoStop}%
\bibitem [{\citenamefont {Decius}(1948)}]{decius:1025}%
  \BibitemOpen
  \bibfield  {author} {\bibinfo {author} {\bibfnamefont {J.~C.}\ \bibnamefont
  {Decius}},\ }\href {\doibase 10.1063/1.1746719} {\bibfield  {journal}
  {\bibinfo  {journal} {The Journal of Chemical Physics}\ }\textbf {\bibinfo
  {volume} {16}},\ \bibinfo {pages} {1025} (\bibinfo {year}
  {1948})}\BibitemShut {NoStop}%
\bibitem [{\citenamefont {Ferigle}\ and\ \citenamefont
  {Meister}(1951)}]{ferigle:982}%
  \BibitemOpen
  \bibfield  {author} {\bibinfo {author} {\bibfnamefont {S.~M.}\ \bibnamefont
  {Ferigle}}\ and\ \bibinfo {author} {\bibfnamefont {A.~G.}\ \bibnamefont
  {Meister}},\ }\href {\doibase 10.1063/1.1748434} {\bibfield  {journal}
  {\bibinfo  {journal} {The Journal of Chemical Physics}\ }\textbf {\bibinfo
  {volume} {19}},\ \bibinfo {pages} {982} (\bibinfo {year} {1951})}\BibitemShut
  {NoStop}%
\bibitem [{\citenamefont {Tannor}(2007)}]{tannor2007iqm}%
  \BibitemOpen
  \bibfield  {author} {\bibinfo {author} {\bibfnamefont {D.}~\bibnamefont
  {Tannor}},\ }\href@noop {} {\emph {\bibinfo {title} {{Introduction to Quantum
  Mechanics: A Time-Dependent Perspective}}}}\ (\bibinfo  {publisher}
  {University Science Books},\ \bibinfo {year} {2007})\BibitemShut {NoStop}%
\bibitem [{\citenamefont {Rheinecker}\ and\ \citenamefont
  {Bowman}(2006)}]{rheinecker2006aic}%
  \BibitemOpen
  \bibfield  {author} {\bibinfo {author} {\bibfnamefont {J.}~\bibnamefont
  {Rheinecker}}\ and\ \bibinfo {author} {\bibfnamefont {J.}~\bibnamefont
  {Bowman}},\ }\href@noop {} {\bibfield  {journal} {\bibinfo  {journal}
  {Journal of Physical Chemistry A}\ }\textbf {\bibinfo {volume} {110}},\
  \bibinfo {pages} {5464} (\bibinfo {year} {2006})}\BibitemShut {NoStop}%
\bibitem [{\citenamefont {Steeves}\ \emph {et~al.}(2008)\citenamefont
  {Steeves}, \citenamefont {Merer}, \citenamefont {Bechtel}, \citenamefont
  {Beck},\ and\ \citenamefont {Field}}]{steeves:1867}%
  \BibitemOpen
  \bibfield  {author} {\bibinfo {author} {\bibfnamefont {A.~H.}\ \bibnamefont
  {Steeves}}, \bibinfo {author} {\bibfnamefont {A.~J.}\ \bibnamefont {Merer}},
  \bibinfo {author} {\bibfnamefont {H.~A.}\ \bibnamefont {Bechtel}}, \bibinfo
  {author} {\bibfnamefont {A.~R.}\ \bibnamefont {Beck}}, \ and\ \bibinfo
  {author} {\bibfnamefont {R.~W.}\ \bibnamefont {Field}},\ }\href@noop {}
  {\bibfield  {journal} {\bibinfo  {journal} {Molecular Physics}\ }\textbf 
  {\bibinfo {volume} {106}}, \bibinfo {pages} {1867} (\bibinfo {year} {2008})}
  \BibitemShut {NoStop}%
\bibitem [{\citenamefont {Tobiason}\ \emph {et~al.}(1993)\citenamefont
  {Tobiason}, \citenamefont {Utz}, \citenamefont {III},\ and\ \citenamefont
  {Crim}}]{tobiason:5762}%
  \BibitemOpen
  \bibfield  {author} {\bibinfo {author} {\bibfnamefont {J.~D.}\ \bibnamefont
  {Tobiason}}, \bibinfo {author} {\bibfnamefont {A.~L.}\ \bibnamefont {Utz}},
  \bibinfo {author} {\bibfnamefont {E.~L.~S.}\ \bibnamefont {III}}, \ and\
  \bibinfo {author} {\bibfnamefont {F.~F.}\ \bibnamefont {Crim}},\ }\href
  {\doibase 10.1063/1.465927} {\bibfield  {journal} {\bibinfo  {journal} {The
  Journal of Chemical Physics}\ }\textbf {\bibinfo {volume} {99}},\ \bibinfo
  {pages} {5762} (\bibinfo {year} {1993})}\BibitemShut {NoStop}%
\bibitem [{\citenamefont {Huet}, \citenamefont {Godefroid},\ and\ \citenamefont
  {Herman}(1990)}]{Huet199032}%
  \BibitemOpen
  \bibfield  {author} {\bibinfo {author} {\bibfnamefont {T.~R.}\ \bibnamefont
  {Huet}}, \bibinfo {author} {\bibfnamefont {M.}~\bibnamefont {Godefroid}}, \
  and\ \bibinfo {author} {\bibfnamefont {M.}~\bibnamefont {Herman}},\ }\href
  {\doibase DOI: 10.1016/0022-2852(90)90306-B} {\bibfield  {journal} {\bibinfo
  {journal} {Journal of Molecular Spectroscopy}\ }\textbf {\bibinfo {volume}
  {144}},\ \bibinfo {pages} {32 } (\bibinfo {year} {1990})}\BibitemShut
  {NoStop}%
\bibitem [{foo({\natexlab{a}})}]{footnote1}%
  \BibitemOpen
  \href@noop {} {\bibinfo {title} {{The consequences of this
  substitution for the kinetic energy will be discussed in {S}ection
  \ref{sec:KE}. This is not a major alteration of the calculation; it
  significantly improves the value obtained for the $v_{2}$ fundamental and
  otherwise causes only minor changes in the level structure. Results quoted
  are solely from this version of the calculation, except in
  Sec.~\ref{sec:cis3161}}}} \BibitemShut {NoStop}%
\bibitem [{\citenamefont {Harris}, \citenamefont {Engerholm},\ and\
  \citenamefont {Gwinn}(1965)}]{harris1965cme}%
  \BibitemOpen
  \bibfield  {author} {\bibinfo {author} {\bibfnamefont {D.}~\bibnamefont
  {Harris}}, \bibinfo {author} {\bibfnamefont {G.}~\bibnamefont {Engerholm}}, \
  and\ \bibinfo {author} {\bibfnamefont {W.}~\bibnamefont {Gwinn}},\
  }\href@noop {} {\bibfield  {journal} {\bibinfo  {journal} {Journal of
  Chemical Physics}\ }\textbf {\bibinfo {volume} {43}},\ \bibinfo {pages}
  {1515} (\bibinfo {year} {1965})}\BibitemShut {NoStop}%
\bibitem [{\citenamefont {Szalay}(1993)}]{szalay:1978}%
  \BibitemOpen
  \bibfield  {author} {\bibinfo {author} {\bibfnamefont {V.}~\bibnamefont
  {Szalay}},\ }\href {\doibase 10.1063/1.465258} {\bibfield  {journal}
  {\bibinfo  {journal} {The Journal of Chemical Physics}\ }\textbf {\bibinfo
  {volume} {99}},\ \bibinfo {pages} {1978} (\bibinfo {year}
  {1993})}\BibitemShut {NoStop}%
\bibitem [{\citenamefont {Marston}\ and\ \citenamefont
  {Balint-Kurti}(1989)}]{marston:3571}%
  \BibitemOpen
  \bibfield  {author} {\bibinfo {author} {\bibfnamefont {C.~C.}\ \bibnamefont
  {Marston}}\ and\ \bibinfo {author} {\bibfnamefont {G.~G.}\ \bibnamefont
  {Balint-Kurti}},\ }\href {\doibase 10.1063/1.456888} {\bibfield  {journal}
  {\bibinfo  {journal} {The Journal of Chemical Physics}\ }\textbf {\bibinfo
  {volume} {91}},\ \bibinfo {pages} {3571} (\bibinfo {year}
  {1989})}\BibitemShut {NoStop}%
\bibitem [{\citenamefont {Colbert}\ and\ \citenamefont
  {Miller}(1992)}]{colbert:1982}%
  \BibitemOpen
  \bibfield  {author} {\bibinfo {author} {\bibfnamefont {D.~T.}\ \bibnamefont
  {Colbert}}\ and\ \bibinfo {author} {\bibfnamefont {W.~H.}\ \bibnamefont
  {Miller}},\ }\href {\doibase 10.1063/1.462100} {\bibfield  {journal}
  {\bibinfo  {journal} {The Journal of Chemical Physics}\ }\textbf {\bibinfo
  {volume} {96}},\ \bibinfo {pages} {1982} (\bibinfo {year}
  {1992})}\BibitemShut {NoStop}%
\bibitem [{\citenamefont {Tuvi}\ and\ \citenamefont {Band}(1997)}]{tuvi:9079}%
  \BibitemOpen
  \bibfield  {author} {\bibinfo {author} {\bibfnamefont {I.}~\bibnamefont
  {Tuvi}}\ and\ \bibinfo {author} {\bibfnamefont {Y.~B.}\ \bibnamefont
  {Band}},\ }\href {\doibase 10.1063/1.475198} {\bibfield  {journal} {\bibinfo
  {journal} {The Journal of Chemical Physics}\ }\textbf {\bibinfo {volume}
  {107}},\ \bibinfo {pages} {9079} (\bibinfo {year} {1997})}\BibitemShut
  {NoStop}%
\bibitem [{\citenamefont {Podolsky}(1928)}]{podolsky}%
  \BibitemOpen
  \bibfield  {author} {\bibinfo {author} {\bibfnamefont {B.}~\bibnamefont
  {Podolsky}},\ }\href {\doibase 10.1103/PhysRev.32.812} {\bibfield  {journal}
  {\bibinfo  {journal} {Phys. Rev.}\ }\textbf {\bibinfo {volume} {32}},\
  \bibinfo {pages} {812} (\bibinfo {year} {1928})}\BibitemShut {NoStop}%
\bibitem [{\citenamefont {Wei}\ and\ \citenamefont {{Carrington,
  Jr.}}(1994)}]{wei:1343}%
  \BibitemOpen
  \bibfield  {author} {\bibinfo {author} {\bibfnamefont {H.}~\bibnamefont
  {Wei}}\ and\ \bibinfo {author} {\bibfnamefont {T.}~\bibnamefont {{Carrington,
  Jr.}}},\ }\href {\doibase 10.1063/1.467827} {\bibfield  {journal} {\bibinfo
  {journal} {The Journal of Chemical Physics}\ }\textbf {\bibinfo {volume}
  {101}},\ \bibinfo {pages} {1343} (\bibinfo {year} {1994})}\BibitemShut
  {NoStop}%
\bibitem [{foo({\natexlab{b}})}]{footnote2}%
  \BibitemOpen
  \href@noop {} {\bibinfo {title} {{This redefinition weights the
  $R_{CC}$ kinetic energy by the projection of the internal coordinate onto the
  $v_{2}$ normal mode. The effect is to discard portions of this term that
  would be included in $v_{1}$ in full dimensionality}}}\BibitemShut {NoStop}%
\bibitem [{\citenamefont {Stanton}\ \emph {et~al.}(2008)\citenamefont
  {Stanton}, \citenamefont {Gauss}, \citenamefont {Watts}, \citenamefont
  {Szalay}, \citenamefont {Bartlett}, \citenamefont {{with contributions from:
  A. A. Auer}}, \citenamefont {Bernholdt}, \citenamefont {Christiansen},
  \citenamefont {Harding}, \citenamefont {Heckert}, \citenamefont {Heun},
  \citenamefont {Huber}, \citenamefont {Jonsson}, \citenamefont {Jus{\`e}lius},
  \citenamefont {Lauderdale}, \citenamefont {Metzroth}, \citenamefont
  {Michauk}, \citenamefont {Price}, \citenamefont {Ruud}, \citenamefont
  {Schiffmann}, \citenamefont {Tajti}, \citenamefont {Varner}, \citenamefont
  {V{\'a}zquez},\ and\ \citenamefont {{(P. R. Taylor), ABACUS (T. Helgaker, H.
  J. Aa. Jensen, P. J{\o}rgensen and J. Olsen)}}}]{cfour}%
  \BibitemOpen
  \bibfield  {author} {\bibinfo {author} {\bibfnamefont {J.~F.}\ \bibnamefont
  {Stanton}}, \bibinfo {author} {\bibfnamefont {J.}~\bibnamefont {Gauss}},
  \bibinfo {author} {\bibfnamefont {J.~D.}\ \bibnamefont {Watts}}, \bibinfo
  {author} {\bibfnamefont {P.~G.}\ \bibnamefont {Szalay}}, \bibinfo {author}
  {\bibfnamefont {R.~J.}\ \bibnamefont {Bartlett}}, \bibinfo {author}
  {\bibnamefont {{with contributions from: A. A. Auer}}}, \bibinfo {author}
  {\bibfnamefont {D.~E.}\ \bibnamefont {Bernholdt}}, \bibinfo {author}
  {\bibfnamefont {O.}~\bibnamefont {Christiansen}}, \bibinfo {author}
  {\bibfnamefont {M.~E.}\ \bibnamefont {Harding}}, \bibinfo {author}
  {\bibfnamefont {M.}~\bibnamefont {Heckert}}, \bibinfo {author} {\bibfnamefont
  {O.}~\bibnamefont {Heun}}, \bibinfo {author} {\bibfnamefont {C.}~\bibnamefont
  {Huber}}, \bibinfo {author} {\bibfnamefont {D.}~\bibnamefont {Jonsson}},
  \bibinfo {author} {\bibfnamefont {J.}~\bibnamefont {Jus{\`e}lius}}, \bibinfo
  {author} {\bibfnamefont {W.~J.}\ \bibnamefont {Lauderdale}}, \bibinfo
  {author} {\bibfnamefont {T.}~\bibnamefont {Metzroth}}, \bibinfo {author}
  {\bibfnamefont {C.}~\bibnamefont {Michauk}}, \bibinfo {author} {\bibfnamefont
  {D.~R.}\ \bibnamefont {Price}}, \bibinfo {author} {\bibfnamefont
  {K.}~\bibnamefont {Ruud}}, \bibinfo {author} {\bibfnamefont {F.}~\bibnamefont
  {Schiffmann}}, \bibinfo {author} {\bibfnamefont {A.}~\bibnamefont {Tajti}},
  \bibinfo {author} {\bibfnamefont {M.~E.}\ \bibnamefont {Varner}}, \bibinfo
  {author} {\bibfnamefont {J.}~\bibnamefont {V{\'a}zquez}}, \ and\ \bibinfo
  {author} {\bibfnamefont {including the integral packages: MOLECULE}~\bibnamefont {{(J. Alml\"of, P. R. Taylor), ABACUS (T.
  Helgaker, H. J. Aa. Jensen, P. J{\o}rgensen and J. Olsen)}}},\ }\href@noop {}
  {\bibinfo {title} {\mbox{CFOUR}},} (\bibinfo {year}
  {2005-2008}).\ \bibinfo {note} {See http://www.cfour.de}\BibitemShut
  {NoStop}%
\bibitem [{\citenamefont {K\'{a}llay}()}]{mrcc}%
  \BibitemOpen
  \bibfield  {author} {\bibinfo {author} {\bibfnamefont {M.}~\bibnamefont
  {K\'{a}llay}},\ }\href@noop {} {\bibinfo {title} {{MRCC, a
  string-based quantum chemical program suite}}.} \bibinfo {note} {See also
  M.~K\'{a}llay, P.~R.~Surj{\'a}n, J. Chem. Phys. 115 2945 (2001) as well as:
  www.mrcc.hu.}\BibitemShut {Stop}%
\bibitem [{\citenamefont {Watson}\ \emph {et~al.}(1982)\citenamefont {Watson},
  \citenamefont {Herman}, \citenamefont {Craen},\ and\ \citenamefont
  {Colin}}]{Watson1982101}%
  \BibitemOpen
  \bibfield  {author} {\bibinfo {author} {\bibfnamefont {J.~K.~G.}\
  \bibnamefont {Watson}}, \bibinfo {author} {\bibfnamefont {M.}~\bibnamefont
  {Herman}}, \bibinfo {author} {\bibfnamefont {J.~C.~V.}\ \bibnamefont
  {Craen}}, \ and\ \bibinfo {author} {\bibfnamefont {R.}~\bibnamefont
  {Colin}},\ }\href {\doibase DOI: 10.1016/0022-2852(82)90242-9} {\bibfield
  {journal} {\bibinfo  {journal} {Journal of Molecular Spectroscopy}\ }\textbf
  {\bibinfo {volume} {95}},\ \bibinfo {pages} {101 } (\bibinfo {year}
  {1982})}\BibitemShut {NoStop}%
\bibitem [{\citenamefont {Lundberg}(1995)}]{lundberg1995CNPI}%
  \BibitemOpen
  \bibfield  {author} {\bibinfo {author} {\bibfnamefont {J.}~\bibnamefont
  {Lundberg}},\ }in\ \href@noop {} {\emph {\bibinfo {booktitle} {{Molecular
  Dynamics and Spectroscopy by Stimulated Emission Pumping}}}},\ \bibinfo
  {editor} {edited by\ \bibinfo {editor} {\bibfnamefont {H.}~\bibnamefont
  {Dai}}\ and\ \bibinfo {editor} {\bibfnamefont {R.}~\bibnamefont {Field}}}\
  (\bibinfo  {publisher} {World Scientific Pub. Co. Inc.},\ \bibinfo {year}
  {1995})\ Chap.~\bibinfo {chapter} {22}, p.\ \bibinfo {pages}
  {791}\BibitemShut {NoStop}%
\bibitem [{\citenamefont {Mordaunt}\ and\ \citenamefont
  {Ashfold}(1994)}]{mordaunt:2630}%
  \BibitemOpen
  \bibfield  {author} {\bibinfo {author} {\bibfnamefont {D.~H.}\ \bibnamefont
  {Mordaunt}}\ and\ \bibinfo {author} {\bibfnamefont {M.~N.~R.}\ \bibnamefont
  {Ashfold}},\ }\href {\doibase 10.1063/1.467635} {\bibfield  {journal}
  {\bibinfo  {journal} {The Journal of Chemical Physics}\ }\textbf {\bibinfo
  {volume} {101}},\ \bibinfo {pages} {2630} (\bibinfo {year}
  {1994})}\BibitemShut {NoStop}%
\bibitem [{\citenamefont {Merer}\ \emph {et~al.}(2003)\citenamefont {Merer},
  \citenamefont {Yamakita}, \citenamefont {Tsuchiya}, \citenamefont {Stanton},
  \citenamefont {Duan},\ and\ \citenamefont {Field}}]{merer2003new}%
  \BibitemOpen
  \bibfield  {author} {\bibinfo {author} {\bibfnamefont {A.}~\bibnamefont
  {Merer}}, \bibinfo {author} {\bibfnamefont {N.}~\bibnamefont {Yamakita}},
  \bibinfo {author} {\bibfnamefont {S.}~\bibnamefont {Tsuchiya}}, \bibinfo
  {author} {\bibfnamefont {J.}~\bibnamefont {Stanton}}, \bibinfo {author}
  {\bibfnamefont {Z.}~\bibnamefont {Duan}}, \ and\ \bibinfo {author}
  {\bibfnamefont {R.}~\bibnamefont {Field}},\ }\href@noop {} {\bibfield
  {journal} {\bibinfo  {journal} {Molecular Physics}\ }\textbf {\bibinfo
  {volume} {101}},\ \bibinfo {pages} {663} (\bibinfo {year}
  {2003})}\BibitemShut {NoStop}%
\bibitem [{\citenamefont {Steeves}(2009)}]{steevesthesis}%
  \BibitemOpen
  \bibfield  {author} {\bibinfo {author} {\bibfnamefont {A.~H.}\ \bibnamefont
  {Steeves}},\ }\emph {\bibinfo {title} {{Electronic Signatures of Large
  Amplitude Motions}}},\ \href@noop {} {Ph.D. thesis},\ \bibinfo  {school}
  {Massachusetts Institute of Technology} (\bibinfo {year} {2009})\BibitemShut
  {NoStop}%
\bibitem [{\citenamefont {Tajti}\ \emph {et~al.}(2004)\citenamefont {Tajti},
  \citenamefont {Szalay}, \citenamefont {Cs\'{a}sz\'{a}r}, \citenamefont
  {K\'{a}llay}, \citenamefont {Gauss}, \citenamefont {Valeev}, \citenamefont
  {Flowers}, \citenamefont {V\'{a}zquez},\ and\ \citenamefont
  {Stanton}}]{tajti:11599}%
  \BibitemOpen
  \bibfield  {author} {\bibinfo {author} {\bibfnamefont {A.}~\bibnamefont
  {Tajti}}, \bibinfo {author} {\bibfnamefont {P.~G.}\ \bibnamefont {Szalay}},
  \bibinfo {author} {\bibfnamefont {A.~G.}\ \bibnamefont {Cs\'{a}sz\'{a}r}},
  \bibinfo {author} {\bibfnamefont {M.}~\bibnamefont {K\'{a}llay}}, \bibinfo
  {author} {\bibfnamefont {J.}~\bibnamefont {Gauss}}, \bibinfo {author}
  {\bibfnamefont {E.~F.}\ \bibnamefont {Valeev}}, \bibinfo {author}
  {\bibfnamefont {B.~A.}\ \bibnamefont {Flowers}}, \bibinfo {author}
  {\bibfnamefont {J.}~\bibnamefont {V\'{a}zquez}}, \ and\ \bibinfo {author}
  {\bibfnamefont {J.~F.}\ \bibnamefont {Stanton}},\ }\href {\doibase
  10.1063/1.1811608} {\bibfield  {journal} {\bibinfo  {journal} {The Journal of
  Chemical Physics}\ }\textbf {\bibinfo {volume} {121}},\ \bibinfo {pages}
  {11599} (\bibinfo {year} {2004})}\BibitemShut {NoStop}%
\bibitem [{\citenamefont {K\'{a}llay}\ and\ \citenamefont
  {Gauss}(2004)}]{kallay:9257}%
  \BibitemOpen
  \bibfield  {author} {\bibinfo {author} {\bibfnamefont {M.}~\bibnamefont
  {K\'{a}llay}}\ and\ \bibinfo {author} {\bibfnamefont {J.}~\bibnamefont
  {Gauss}},\ }\href {\doibase 10.1063/1.1805494} {\bibfield  {journal}
  {\bibinfo  {journal} {The Journal of Chemical Physics}\ }\textbf {\bibinfo
  {volume} {121}},\ \bibinfo {pages} {9257} (\bibinfo {year}
  {2004})}\BibitemShut {NoStop}%
\bibitem [{foo({\natexlab{c}})}]{footnote3}%
  \BibitemOpen
  \href@noop {} {\bibinfo {title} {{The results in
  Sec.~\ref{sec:cis3161} are based on the first version of the calculation
  mentioned in Sec.~\ref{sec:PES}. The zero-order energy ranking of the levels
  makes it impossible for the second version of the calculation to capture this
  interaction correctly, since the interacting levels cannot cross. In the
  first version of the calculation, the \emph{trans} $v_{3}$ fundamental is
  1058.8 cm$^{-1}$ (as opposed to 1033.6 cm$^{-1}$), such that \emph{trans}
  $3^{4}$ lies considerably higher, and crucially, above rather than below
  \emph{cis} $3^{1}6^{1}$ in zero-order}}}\BibitemShut
  {NoStop}%
\bibitem [{foo({\natexlab{d}})}]{footnote4}%
  \BibitemOpen
  \href@noop {} {\bibinfo {title} {{The reduced dimension potential
  does not allow the CH bond distances to be different, yet their asymmetry is
  appreciable (See Table \ref{tab:TSabinitio}). Relaxing these bond distances
  would give a lower, more accurate barrier height}}}\BibitemShut {NoStop}%
\bibitem [{\citenamefont {Jacobson}, \citenamefont {Silbey},\ and\
  \citenamefont {Field}(1999)}]{jacobson:845}%
  \BibitemOpen
  \bibfield  {author} {\bibinfo {author} {\bibfnamefont {M.~P.}\ \bibnamefont
  {Jacobson}}, \bibinfo {author} {\bibfnamefont {R.~J.}\ \bibnamefont
  {Silbey}}, \ and\ \bibinfo {author} {\bibfnamefont {R.~W.}\ \bibnamefont
  {Field}},\ }\href {\doibase 10.1063/1.478052} {\bibfield  {journal} {\bibinfo
   {journal} {The Journal of Chemical Physics}\ }\textbf {\bibinfo {volume}
  {110}},\ \bibinfo {pages} {845} (\bibinfo {year} {1999})}\BibitemShut
  {NoStop}%
\end{thebibliography}%

\end{document}